\documentstyle[preprint,epsf,aps]{revtex}
\newcommand{\be}{\begin{equation}}
\newcommand{\ee}{\end{equation}}
\newcommand{\ba}{\begin{array}}
\newcommand{\ea}{\end{array}}
\newcommand{\bea}{\begin{eqnarray}}
\newcommand{\eea}{\end{eqnarray}}
\newcommand{\bit}{\begin{itemize}}
\newcommand{\eit}{\end{itemize}}
\newcommand{\ben}{\begin{enumerate}}
\newcommand{\een}{\end{enumerate}}
\newcommand{\tm}{\item}
\newcommand{\up}{\uparrow}
\newcommand{\down}{\downarrow}
\newcommand{\btm}{\bibitem}

\begin{document}

\title{\large \bf  One-, Two-, and Three-channel Kondo effects \\
for a model Ce$^{3+}$ Impurity in a Metal}
\author{Tae-Suk Kim and D. L. Cox}
\address{Department of Physics, Ohio State University, Columbus, Ohio 43210,
USA}
\date{\today}
\maketitle
\begin{abstract}
 We present studies of a simple Anderson model Hamiltonian for
Ce$^{3+}$ ions in cubic symmetry
with three configurations ($f^0$,$f^1$,$f^2$).
Our Hamiltonian includes: (i) One-channel Anderson model; (ii) Two-channel
Anderson model.
Using the third order scaling analysis, we study stability of the non-Fermi
liquid fixed point of the two-channel Kondo model for
Ce$^{3+}$ ions in cubic symmetry against the one-channel Kondo interaction.
Using the non-crossing approximation (NCA),
we also report detailed studies of our simplified
model of one-channel \& two-channel Anderson model
which exhibits competition between the Fermi-liquid fixed point of
the one-channel Kondo model and the non-Fermi fixed point of
the two-channel Kondo model.
We provide the phase diagram in the model
parameter space and study the thermodynamics and the transport properties of
our simplified model Hamiltonian.
Thermodynamics and transport coefficients
show the distinct behaviors for different numbers of channels.
We confirm in detail that the NCA is a valid numerical
method for the overcompensated multichannel
$S_I=1/2$ Anderson models. Our model study might be relevant to the non-Fermi
liquid alloy Ce$_{1-x}$La$_x$Cu$_{2.2}$Si$_2$.

\end{abstract}

\section{Introduction}
 The Kondo effect\cite{kondo}
has been of great interest in condensed matter physics
since its observation. The proposed model Hamiltonian, a magnetic
$S_I=1/2$ local moment interacting with the conduction electron gas,
looked very
simple but was non-trivial due to many body nature of the problem.
Ever since, many generalized models have been studied to extend our
understanding and to relate to real materials.
The simplest $S_I=1/2$ orbitally nondegenerate
Anderson model\cite{anderson} and s-d exchange model are now
well understood for a single impurity case using several techniques.
The numerical renormalization group (NRG)\cite{nrg}
method was able to provide complete information about crossover
from the high temperature fixed point to the low temperature
fermi liquid fixed point for these models.
Subsequently, the exact diagonalization of these
models was realized by the Bethe Ansatz (BA)\cite{bethe},
which also gives the exact
solution for thermodynamics of these models.
However, it has not proven possible to compute dynamical properties with
the BA.
Through non-crossing approximation (NCA)\cite{nca},
dynamics as well as thermodynamics\cite{bcw}
have been extensively studied for the infinite on-site Coulomb interaction
models.
The Quantum monte Carlo method (QMC)\cite{qmc}
has also been applied
to study statics and dynamics for the simple $S_I=1/2$ models.
Recently conformal field theory (CFT)\cite{cft,ludwig}
has been used to study all properties near
the low temperature fixed points.
However we are still far from a complete understanding for realistic
models which, for example, include the strong spin-orbit coupling, the
crystal electric field effects, and multiple (more than two) configurations.
In this paper, we study a realistic extension of the conventional simple
approach to the Kondo effect for Ce$^{3+}$ ions,
including the strong spin-orbit coupling,
the crystal electric field effect, and multiple configurations
in their simplest form.

 Ever since Nozi\`{e}res and Blandin\cite{nozbland} introduced the
multichannel Kondo model, its realization in real materials
has been controversial.
On the theoretical side, the multichannel Kondo models
are well understood\cite{cft,twoch1,twoch2,sacramento}
irrespective of the experimental situations.
When the degenerate channel number ($N_{\rm ch}$), which is the number of the
conduction electron ``bands" coupled to the impurity site, is greater than
$2S_I$ (the impurity spin), the impurity magnetic moment is
overscreened and the strong coupling fixed point (Fermi liquid) becomes
unstable leading to a non-trivial fixed point (non-Fermi liquid behavior).
In the underscreened and the completely screened cases,
$N_{\rm ch} \leq 2S_I$, the strong coupling fixed point is stable
resulting in a Fermi liquid ground state.
The spin susceptibility $\chi (T)$ and
specific heat coefficient $ C(T) / T$ for the
two-channel $S_I=1/2$ magnetic Kondo model
are proportional to $\log (T_K/T)$ at low temperatures
\cite{cft,ludwig,twoch1,twoch2,sacramento}, where
$T_K$ is the Kondo energy scale.
The dynamic susceptibility shows Marginal Fermi liquid behavior
\cite{coxruck}. The resistivity increases logarithmically and saturates to
a constant with $\rho (T) = \rho (0) ~[~ 1 - a \sqrt{T/T_{\rm K}} ~]$ with
decreasing temperature\cite{cft}.
On the other hand, the one-channel $S_I=1/2$
Kondo model leads to the Fermi liquid ground state.
In that case, magnetic susceptibility $\chi (T)$ and
specific heat coefficient $ C(T) / T$ saturate to constants
of order $1/T_K$\cite{nrg,bethe}.
Resistivity increases logarithmically and saturates to
a constant with $\rho (T) = \rho (0) ~[~ 1 - a (T/T_{\rm K})^2 ~]$ with
decreasing temperature\cite{noz}.

 In this paper we study
a model Hamiltonian for Ce$^{3+}$ ions in cubic
metals with three configurations ($f^0$, $f^1$, $f^2$).
The nominal ground configuration $f^1$
can fluctuate to $f^0$ and $f^2$ configurations by hybridizing with
the conduction electrons. A one-channel Anderson hybridization interaction
is present between $f^0$ and $f^1$ configurations. A two-channel Anderson
hybridization interaction
is present between $f^1$ and $f^2$ configurations.
We report detailed studies of our simplified Hamiltonian
of one-channel \& two-channel Anderson model using the NCA.
This simple model is quite intriguing in that we can study
the competition between the two different Kondo effects, that is,
the Fermi liquid and the non-Fermi liquid fixed points.
The distinct ground state physics for the different numbers of channels
is classified using the zero temperature analysis of NCA integral equations
and the third order scaling theory. We calculate the thermodynamics and the
dynamics of this simple model and find that all the calculated physical
quantities show the behavior appropriate for the accessible different
channel numbers. The static magnetic
susceptibility displays a scaling behavior agreeing with the exact Bethe
ansatz results in the two- and three-channel cases. NCA calculation of
entropy and specific heat is also compared with the Bethe ansatz results.
The resistivity shows the correct temperature dependence near zero temperature
agreeing with the conformal field theory results in the two- and
three-channel cases. The sign and the magnitude of the thermopower are
dependent sensitively on the relevant channel numbers. The peak position
in the dynamic magnetic susceptibility is almost linear in temperature
in the overscreened cases. A short paper which presents some of these
results will appear elsewhere\cite{kimcox}.

 Our study is motivated by a recent discovery of non-Fermi liquid system,
Ce$_x$La$_{1-x}$Cu$_{2.2}$Si$_2$ ($x=0.1$)\cite{andraka}.
We summarize the experimental findings of this alloy system.
The logarithmic divergence in both the magnetic
susceptibility $\chi(T)$ and the specific heat linear coefficient
$\gamma (T)$  have been observed for
Ce$_x$La$_{1-x}$Cu$_{2.2}$Si$_2$ ($x=0.1$).
The two-channel $S_I=1/2$ magnetic Kondo physics
\cite{cft,ludwig,twoch1,twoch2,sacramento} provides a theoretical framework
to explain the thermodynamics of this system at low temperatures.
$\gamma (T)$ initially increases in the presence of the magnetic field,
which qualitatively agrees with the two-channel Kondo effect coming from
the lifting of residual entropy\cite{sacramento}.
In the one-channel Kondo effect, the
Sommerfeld coefficient decreases in the magnetic field due to
the destruction of the Kondo effect.
The Wilson ratio is estimated to be $R \approx 2.7$ from the slopes of two
curves ($\chi (T)$ and $\gamma (T)$), which compares well with
theoretical value $8/3$ for the two-channel magnetic $S_I=1/2$ Kondo
model\cite{ludwig}.
The good agreement between the theoretical and the experimental Wilson
ratios supports our crystalline electric field energy scheme described
below. This system is pseudo-cubic (i.e., the crystal field scheme on the
Ce$^{3+}$ site appears cubic). The best superconducting
system with excess Cu shows almost isotropic magnetic
susceptibility\cite{chitep}.
That the pseudo-cubic $\Gamma_7$ magnetic doublet in $f^1$ lies lowest
is inferred from
the neutron scattering experiments\cite{neutron}.
The thermopower for CeCu$_2$Si$_2$
changes sign near $70$ K and stays negative below with a large
extremum ($-20~\mbox{to}~-30 \mu$V/K)\cite{chitep},
suggesting the presence of
strong hole resonance scattering.
As we will show below, these thermopower
results also support our interpretation of the two channel magnetic
Kondo physics. Though other experiments (e.g., specific heat and magnetic
susceptibility) support the interpretation of them in terms of the
two-channel Kondo effect, a linear temperature dependence in the resistivity
remains as a puzzle.
The $\sqrt{T}$ behavior in the resistivity
is predicted from conformal field theory treatment of the two-channel
Kondo models\cite{cft,ludwig}. In Fig.~\ref{resfit}, we present our
numerical calculation of resistivity and the experimental results
measured in the alloy system Ce$_x$La$_{1-x}$Cu$_{2.2}$Si$_2$ ($x=0.1$)
\cite{andraka}. It can be seen that the data curves downwards at lower
temperatures which may indicate a crossover to a new fixed point.
{}From all these experimental findings, we believe
that the alloy system
Ce$_x$La$_{1-x}$Cu$_{2.2}$Si$_2$ ($x=0.1$) is a strong candidate
for two-channel $S_I=1/2$ magnetic Kondo system.

 In addition, the two-channel, $S_I=1/2$ Kondo effect may be realized
in other materials: two-channel quadrupolar Kondo effect\cite{quad}
in some U alloys and two-level systems\cite{tls} in metallic point contacts.
U alloy systems include U$_{0.2}$Y$_{0.8}$Pd$_3$
\cite{uypd}, U$_x$Th$_{1-x}$Ru$_2$Si$_2$ \cite{urusi},
UCu$_{3.5}$Pd$_{1.5}$ \cite{ucupd},
U$_{0.1}$Th$_{0.9}$Ni$_2$Al$_3$ \cite{unial},
U$_{0.1}$Pr$_{0.9}$Ni$_2$Al$_3$ \cite{unial}, U$_x$Sc$_{1-x}$Pd$_3$
\cite{uscpd}, U$_{0.9}$Th$_{0.1}$Be$_{13}$\cite{ube13},
and U$_{x}$Th$_{1-x}$Pd$_2$Al$_3$\cite{updal}.
All the above systems show a logarithmic divergence at low temperature
in the linear specific heat coefficient and a square root saturation in the
static magnetic susceptibility. U$_x$Th$_{1-x}$Ru$_2$Si$_2$ \cite{urusi}
is an exception showing a logarithmically divergent magnetic susceptibility.
The two channel quadrupolar Kondo physics has been invoked to
explain non-Fermi liquid behavior in thermodynamic and
transport properties of U$_x$Y$_{1-x}$Pd$_3$ for $x=0.2$ \cite{uypd} and
other U alloy systems. Recently, the resistivity in a metallic constriction
was observed to obey $\sqrt{T}$ behavior and was interpreted as due to
two-channel Kondo scattering from atomic two-level tunneling systems
\cite{cornel,ufl}.

 Our paper is organized as follows.
In section II, we introduce our simple model Hamiltonian and analyze
this model using the third order scaling equation.
We briefly introduce the NCA in section III.
Zero temperature analysis of NCA integral equations follows in section IV.
In section V, we present the detailed numerical analysis of our simple model
Hamiltonian. We conclude in the section VI.

\section{Model Hamiltonian}
 The single impurity Anderson model\cite{anderson} has been very successful
\cite{bcw}
in describing Kondo systems (meaning magnetic transition metal elements
embedded in normal metals, and dilute rare earth or actinide alloys).
The thermodynamics is rather well explained by the single impurity
properties for even highly concentrated Ce alloys\cite{bcw,singleion}.
Coherence effects, arising from the lattice of Anderson or
Kondo ions at low temperatures, do not play such an important role
in thermodynamics. Transport properties also are well explained by
the single impurity model except for the low temperature regime
where a coherence effect prevails due to the coherent Bloch state
formation leading to vanishing resistivity at zero temperature
(the residual resistivity is larger than the room temperature one
in the dilute impurity limit).
This one-channel Anderson model can explain the complete screening of the
magnetic moment at the local moment sites leading to
the local Fermi liquid ground state discussed in the introduction.

 The two-channel orbital Kondo model (quadrupolar or two-level system)
keeps the channel symmetry guaranteed by time reversal symmetry, but
can suffer from the ground state degeneracy lifting due to the Jahn-Teller
effect.
In general, the orbital Kondo model has the exchange anisotropy.
It has been shown that the exchange anisotropy is irrelevant for
the two-channel $S_I=1/2$ Kondo models\cite{anisoexchange}.
As will be shown in other publication\cite{kimcoxtobe},
the two-channel magnetic Kondo model
for Ce$^{3+}$ is vulnerable to the channel asymmetry due to its orbital
nature in the channel degrees of freedom\cite{nozbland}. However,
the channel-mixing Kondo interaction can save the two-channel physics
\cite{kimcoxtobe}.

 The conventional theory of Ce$^{3+}$ impurities assumes the infinite
on-site Coulomb interaction, which removes the $f^2$ configuration
from consideration\cite{bcw}, and
as a result has no chance to get the two-channel Kondo effect
which we shall explain below. When we relax our assumption about the
strong on-site Coulomb interaction and we include the detailed
atomic energy structure, we can develop a variety of
model Hamiltonians\cite{kimcoxtobe}.

 In our simple model, we assume the magnetic $f^1~J=5/2~\Gamma_7$ CEF
doublet lies lowest in the $f^1$ configuration, and we keep the two
excited states -- a singlet $f^0$ and the nonmagnetic $f^2 ~ \Gamma_3$
CEF doublet.
We find the one-channel Anderson model in mixing between $f^0$ and $f^1$
configurations and the two-channel Anderson model in mixing
between $f^1$ and $f^2$ configurations\cite{coxham}.
Other interesting Kondo interactions\cite{kimcoxtobe} arise
when the excited triplets in the $f^2$
configuration are included.
According to the group theory analysis, the hybridization is mediated
only by the cubic $\Gamma_8$ conduction electrons between $f^1$ and $f^2$
($\Gamma_3 \otimes \Gamma_7 = \Gamma_8$),
and by $\Gamma_7$ between $f^0$ and $f^1$ for the mixing potential
allowed in the cubic crystal. CEF states are schematically drawn in
the Fig.~\ref{cef} for this simple model.
To see the essential physics, we restrict our attention to
the simple case of isotropic hybridization and a free conduction band
with Lorentzian/Gaussian density of states (DOS). In this simple case,
two components of conduction partial wave $j_c =5/2, 7/2$ can mix with
the atomic orbitals.
For the moment, we will consider only one angular momentum component,
say, $j_c =5/2$ of the conduction band for our model study.
We shall examine the effects of relaxing this in the other paper
\cite{kimcoxtobe}.
Our model Hamiltonian is
\bea
 H &=& H_{\rm cb} + H_{\rm at} + H_1, \\
 H_{\rm cb}
  &=& \sum_{\epsilon  \alpha} \epsilon
      c_{\epsilon \Gamma_7\alpha}^{\dagger}  c_{\epsilon \Gamma_7\alpha}
   + \sum_{\epsilon n \alpha} \epsilon
      c_{\epsilon \Gamma_8n \alpha}^{\dagger}
      c_{\epsilon \Gamma_8n \alpha}, \\
 H_{\rm at}
  &=& \epsilon_0 |f^0><f^0| + \epsilon_1 \sum_{\alpha} |f^1 \Gamma_7\alpha >
     < f^1 \Gamma_7\alpha |
   + \epsilon_2 \sum_{n}|f^2 \Gamma_3n >< f^2 \Gamma_3n |,
      \\
 H_1 &=& V_{01} \sum_{\epsilon \alpha} c_{\epsilon \Gamma_7\alpha}^{\dagger}
      | f^0 >< f^1 \Gamma_7\alpha | + h.c. \nonumber\\
  && + V_{12} \sum_{\epsilon n \alpha} (-1)^{\alpha +1/2}
     c_{\epsilon \Gamma_8n \alpha}^{\dagger}
           |f^1 \Gamma_7\bar{\alpha} > < f^2 \Gamma_3 n | + h.c.
\eea
$c_{\epsilon\Gamma_7\alpha} (c_{\epsilon\Gamma_7\alpha}^{\dagger})$
and $c_{\epsilon \Gamma_8n \alpha} (c_{\epsilon \Gamma_8n \alpha}^{\dagger})$
are the annihilation (creation)
operators for conduction electrons of $\Gamma_7$ and $\Gamma_8$,
respectively. $\alpha = \up,\down $ denotes the time reversal
pair for the irreducible $\Gamma_7$ representation, and
$n = \pm$ is the quadrupolar index for $\Gamma_3$ and here acting
as the channel indices. The conduction electrons
are assumed to be described by the uncorrelated Lorentzian/Gaussian
density of states with bandwidth $D=3$ eV.
$\epsilon_{0,1,2}$ are the configuration energies for the
empty ($\epsilon_0 = 0$), single, and double occupancies, respectively.
We lumped all Clebsch-Gordon
coefficients into the hybridization constants, $V_{01}$ and $V_{12}$,
except for the phase dependence on the cubic degeneracy indices.
The phase dependence on the Kramers doublet index comes from time
reversal symmetry.
We will consider these two hybridization constants to be independent of
each other to probe the competition between the one-channel
and the two-channel Kondo physics.

 When the real charge fluctuations are removed from the model system
in the Kondo limit, we have to construct the tensor operators representing
each CEF states for the $f^1$ configuration and the projected conduction
electron CEF states. In this paper, the relevant tensor operators are
for $\Gamma_{7,8}$ CEF states. We can show using the character table
\cite{group} that
\bea
\Gamma_7 \otimes \Gamma_7
 &=& \Gamma_1 \oplus \Gamma_4, \\
\Gamma_8 \otimes \Gamma_8
 &=& \Gamma_1 \oplus \Gamma_2 \oplus \Gamma_3 \oplus
    2 \Gamma_4 \oplus 2\Gamma_5.
\eea
In the direct product, the first CEF states are written as ket, and the
second as bra. The $\Gamma_7$ tensor operator ($2\times 2$ tensor)
is a direct sum of charge operator ($\Gamma_1$) and
spin operator ($\Gamma_4$). Indeed,
the Schrieffer-Wolff transformation leads to two interaction
terms: the spin exchange
interaction and the pure potential scattering term.
The relevant terms of $\Gamma_8$ tensor operators are $2\Gamma_4$ in our
model. In the conduction electron $\Gamma_8$ tensor space, one of the two
$\Gamma_4$'s gives rise to the ordinary $S_c=1/2$ spin operators with two
degenerate orbital channels
and the other to the $S_c=3/2$ spin operator with one channel\cite{kimcoxtobe}.

 Hence, there are three distinct channel labels for conduction electron partial
wave states about the Ce impurity in this simple model.
One channel is just the $\Gamma_7$
doublet. The other two are the $\Gamma_{3}\pm$ ``orbital" states of the
$\Gamma_8$ quartet. Each $\Gamma_{3}\pm$ orbital has a $\Gamma_7$
``spin" doublet (recall $\Gamma_8 = \Gamma_7 \otimes \Gamma_3$.)
As shown in Fig.~\ref{channel} for $J=5/2$ conduction partial waves,
the $+$ orbital is ``stretched" along
the quantized axis (one of the three principal cubic axes, taken to be
$\hat{z}$ here for definiteness.)
The ``$-$" orbital is ``squashed" in the $xy$ plane.
We note that the
simplest example of $\Gamma_8$ partial wave quartet is, for zero spin-orbit
coupling, $d$-wave states with $+ \to 3z^2 - r^2$ and $- \to x^2-y^2$.
The ``spin" index is then real spin of the electrons.

 In the Kondo limit with the stable $f^1$ configuration, we may
remove the $f^0-f^1, f^1-f^2$ charge fluctuations from the Hamiltonian of
Eq. (4) using the Schrieffer-Wolff
transformation \cite{sw} to find the effective Hamiltonian.
\bea
\tilde{H}_1
 &=&  J_1 \vec{S}_{c\Gamma_7} (0) \cdot \vec{S}_{\Gamma_7}
  + J_2 \sum_{n=\pm} \vec{S}_{c\Gamma_8 n} (0) \cdot
            \vec{S}_{\Gamma_7} \\
J_1 &=& { 2|V_{01}|^2 \over - \epsilon_1 }, ~~
 J_2 = { 2|V_{12}|^2 \over \epsilon_2 - \epsilon_1 }, \\
\vec{S}_{\Gamma_7}
 &=& { 1\over 2} \sum_{\alpha\beta}
   |f^1;\Gamma_7\alpha > \vec{\sigma}_{\alpha\beta} <f^1;\Gamma_7\beta |.
\eea
$\vec{S}_{\Gamma_7}$ is the $f^1$ pseudo spin.
$\vec{S}_{c\Gamma_7} (0)$ and $\vec{S}_{c\Gamma_8 \pm} (0)$
are the conduction electron pseudo spin densities at the impurity
site of symmetry $\Gamma_7$ and $\Gamma_8$, respectively.
When the $f^1$ configuration is stable, its pseudo spin is coupled to
the conduction band in a one-channel via $f^0$ configuration,
and in a two-channel via $f^2$ configuration.
Their couplings are both antiferromagnetic.
 The unique feature of our Hamiltonian is that it can generate
the one-, two-, and three-channel ground states depending on
the model parameters. The competition between the Fermi liquid fixed point
and the non-Fermi liquid fixed point can thus be investigated
using this model Hamiltonian.

 We first analyze our simple model Hamiltonian using the third
order scaling argument, i.e., the perturbative renormalization group (RG)
\cite{nozbland,scale}.
At temperature $T$, only the conduction electrons (thermally excited)
inside the band of order $T$ with respect to the Fermi level play an
important role in determining physical properties.
Thus we can integrate out the band edge states (virtually excited states)
to find the effective Hamiltonian. Though the following analysis is
restricted to $\omega \ll D$(the conduction band width)
and the perturbative regime (weak coupling limit),
we can derive qualitative results out of this. For quantitative results,
a full numerical renormalization group (NRG) study is required.

 It can be deduced from the scaling theory that the low
temperature and the low energy physics is dominated by the one-channel
or the two-channel Kondo effects depending on their relative
magnitude of the antiferromagnetic couplings.
We introduce an exchange coupling matrix in the channel space which
is convenient for the derivation of the scaling equations.
We can thus rewrite the one-channel and two-channel Kondo model
in the following form\cite{kimcoxtobe}.
\bea
\tilde{H}_1
 &=& {\bf J} \otimes \vec{S}_c (0) \cdot \vec{S}_I, \\
{\bf J} &=& \pmatrix{J_1 & 0 & 0 \cr
               0 & J_2 & 0 \cr 0 & 0 & J_2}.
\eea
Here $\vec{S}_c$ and $\vec{S}_I$ are $S=1/2$ operators.
The scaling equations
of our simple model Hamiltonian up to the third order diagrams of
Fig.~\ref{scale} are
\bea
{\partial {\bf g} \over \partial x}
 &=& {\bf g}^2 - {1\over 2} ~ {\bf g} ~ \mbox{Tr} [{\bf g}^2],
   \\
{\bf g} &=& N(0) {\bf J}.
\eea
The scaling equations in components are
\bea
{\partial g_1 \over \partial x}
 &=& g_1^2 - {1\over 2} ~g_1~[~g_1^2 + 2 g_2^2 ~], ~~ g_1 = N(0)J_1 > 0, \\
{\partial g_2 \over \partial x}
 &=& g_2^2 - {1\over 2} ~g_2~[~g_1^2 + 2 g_2^2 ~], ~~ g_2 = N(0)J_2 > 0.
\eea
Here $x = \log(D/T)$. We can identify three fixed points related
to one-, two-, and three-channel Kondo physics.
The one-channel, strong coupling
fixed point $(g_1^*, g_2^*) = (\infty, 0)$ is stable leading
to the Fermi liquid ground state\cite{scale}.
The three-channel fixed point $(2/3, 2/3)$ is
stable along the line $g_1 = g_2$ in the $g_1-g_2$ plane, but unstable for
any small perturbation from $g_1 = g_2$.
Finally, the two-channel fixed point
$(0, 1)$ is stable leading to the logarithmically divergent thermodynamic
properties at zero temperature.
{}From the scaling analysis, we can infer the ground state physics:
one-channel for $J_1 > J_2$; two-channel for $J_1 < J_2$;
and three-channel for $J_1 = J_2$. As will be shown in
section IV, the zero temperature analysis of the NCA equations leads to
the same conclusion.

 We now discuss the neglected other 8 $\Gamma_3$ irreps in the
$f^2$ configuration\cite{kimcoxtobe}.
The 9 $\Gamma_3$ CEF states all contribute to the enhancement of the
two-channel exchange coupling between $f^1 \Gamma_7$ spin and the $\Gamma_8$
conduction electron spins.
\bea
H_1 &=& \sum_{\epsilon} \sum_{i n\alpha} (-1)^{\alpha +1/2} V_{12}^i ~
    c_{\epsilon \Gamma_8 n \alpha}^{\dagger} ~
    |f^1 \Gamma_7\bar{\alpha} > <f^2 \Gamma_3^i n|
     + h.c.
\eea
The NCA can treat this problem with
the extension that now the $f^2 ~ \Gamma_3$ Green's function becomes
an $9\times 9$ matrix. See the Appendix for more details.
The Schrieffer-Wolff transformation leads to
\bea
\tilde{H}_1
 &=& J ~ \sum_{n=\pm} \vec{S}_{c\Gamma_8 n} \cdot \vec{S}_{\Gamma_7}, \\
J &=& \sum_{i=1}^{11} ~ {2 |V_{12}^{i}|^2 \over \epsilon_2^{i} - \epsilon_1}.
\eea
Here $\epsilon_2^{i}$ is the energy level for the $i$-th $f^2 \Gamma_3$
state. Hence multiple $\Gamma_3$ states in the $f^2$ configuration
lead to the enhancement of the two-channel exchange coupling.

 Particle-hole asymmetry in the conduction band density of states (DOS)
is also important in determining the ground state weights of the
one-channel ($f^0-f^1$) and the two-channel ($f^1-f^2$)
contributions. In the scaling approach, the particle-hole
asymmetry is completely neglected. However the NCA can
take into account this particle-hole asymmetry.
The occupied conduction electron states (hole) contribute to the
$f^0$ self energy (see the section III), while the unoccupied
states (particle) to the $f^2$ self energy. Hence the more
weight in the particle side can enhance the effective
hybridization strength between $f^1$ and $f^2$ configurations.
The particle-dominant conduction band DOS will lead to the
enhancement of the two-channel exchange coupling.

\section{Non-crossing approximation (NCA).}
 We now apply the non-crossing approximation (NCA) \cite{nca}
to study our simple model system -- one-channel \& two-channel Anderson
model. Though this model is highly simplified compared with the full
model\cite{kimcoxtobe}, we can simulate the full model using this
simple model Hamiltonian to study the different physics for varying
channel number.

 In the NCA, our starting basis is the conduction band plus
the atomic Hamiltonian
projected to the atomic electron Fock space and treat the hybridization
between the conduction band and the atomic orbital as a perturbation.
The strength of this approach is that the strong on-site Coulomb interaction
for atomic electrons is treated accurately at the outset.
Pseudo particle Green's functions are introduced for each atomic
electron occupation state which is neither fermionic nor bosonic.
The price we pay is that we cannot apply the conventional Feynmann
diagram techniques to this strongly correlated problems. Thus special
Green's function technique was developed by many investigators
\cite{nca,bcw,grewe,kuramoto,coleman,zhanglee}.
This approach may be justified by expansion
in $1/N$ which reorders the diagrams by treating $NV^2$ as ${\cal O}(1)$.
Here $N$ is the nominal atomic ground state degeneracy and $V$ is the
hybridization strength between the conduction band and the atomic orbitals.
In the NCA, pseudo particle self energy diagrams
include only the leading order skeleton (non-crossing) diagrams
and they are solved self-consistently.
For the one-channel models, this theory includes all the diagrams
up to ${\cal O} (1/N)$ order and a subset of higher order diagrams up to
the infinite order. Though vertex corrections, which is ${\cal O} (1/N^2)$,
are not included, this approach is a conserving approximation\cite{kuramoto}.
For the overscreened multi-channel Anderson models, it has been shown
\cite{coxruck} that the $1/N$ approach becomes exact in the limit
$M,N\to \infty$ with fixed $M/N$ ratio ($M$ is the number of channels).

When we study the most general three-configuration model,
the same symmetry conduction electron can be involved in the two mixing
processes, e.g., $f^0-f^1$ and $f^1-f^2$ for Ce$^{3+}$ atoms.
Generally, a specific vertex correction is required to get the right Kondo
energy scales in this case.
Recently, vertex corrections were included in the study of
the finite $U$ spin $1/2$ Anderson model\cite{pruschke}.
A simplifying feature of our model Hamiltonian
is that the leading vertex correction vanishes, since two different
symmetry conduction electrons are involved in the hybridizations
$f^0 - f^1$ ($\Gamma_7$) and $f^1 - f^2$ ($\Gamma_8$).
This feature greatly simplifies the numerical work and formalism.

 From the leading order skeleton diagrams of Fig.~\ref{nca},
we find the self-consistent integral equations:
\bea
\Sigma_0 (z)
 &=& {\Gamma_{01} \over \pi} \sum_{\alpha} \int d\epsilon ~
   \tilde{N} (\epsilon) f(\epsilon) G_1 (z+\epsilon), \\
\Sigma_1 (z)
 &=& {\Gamma_{01} \over \pi} \int d\epsilon ~\tilde{N} (\epsilon)
   f(-\epsilon) G_0 (z-\epsilon) \nonumber\\
 && + {\Gamma_{12} \over \pi}  \sum_{n} \int d\epsilon ~\tilde{N} (\epsilon)
   f(\epsilon) G_2 (z+\epsilon), \\
\Sigma_2 (z)
 &=& {\Gamma_{12} \over \pi}  \sum_{\alpha} \int d\epsilon ~
 \tilde{N} (\epsilon) f(-\epsilon) G_1 (z-\epsilon), \\
G_N (z) &=& {1 \over z - \epsilon_N -\Sigma_N (z) }, ~~
 \Gamma_{ij} \equiv \pi N(0) |V_{ij}|^2.
\eea
Here $\tilde{N} (\epsilon)$ is the normalized conduction band DOS at the
fermi level such that $\tilde{N} (0) =1$. $\Sigma_{0,1,2} (z)$ and
$G_{0,1,2} (z)$ are the self energy equations and Green's functions
for $f^0$, $f^1 \Gamma_7$, and $f^2 \Gamma_3$ atomic states, respectively.
$f(\epsilon)$ is the Fermi-Dirac distribution function. $\Gamma_{ij}$
is the hybridization strength characterizing the width of the renormalized
atomic electron spectral function peak. One of the strong points of the NCA
approach is that we can easily study any form of the conduction band DOS,
as opposed to other approaches, e.g., Bethe ansatz or conformal field theory,
to Anderson/Kondo models.
As a concrete example,
we will use the structureless Lorentzian/Gaussian DOS to see
the many body physics.
We solved these coupled integral equations numerically
to study the thermodynamics and the dynamics of the model Hamiltonian.

 Pseudo particle Green's functions are not directly measurable. All the
physically measurable quantities are given by the convolution of the
pseudo particle Green's functions. Now it is convenient to introduce
the spectral function ($A_N (\omega)$) for each pseudo particle
Green's function and its negative frequency spectral function
($a_N (\omega)$).
\bea
A_N (\omega)
 &\equiv & - { 1 \over \pi}  \hbox{Im}
    { 1 \over \omega - \epsilon_N - \Sigma_N (\omega) }, \\
a_N (\omega)
 &\equiv& e^{-\beta \omega} ~ A_N (\omega).
\eea
Since $a_N (\omega)$ always appears in combination
with the impurity partition function $Z_{\rm f}$, there is arbitraryness
in overall factor in its definition. The above NCA integral equations
do not have $a_N (\omega)$ in them. Whenever
we calculate any measurable physical quantities, the spectral functions
appear with the Boltzman thermal factor divided by the impurity partition
function.
In the numerical work, we calculate $a_N (\omega)$'s self-consistently.
Roughly speaking, $a_N (\omega)$'s are occupancy
distribution function weighted by the thermal factor.

The impurity partition function for our simple model is defined by
\bea
Z_{\rm f}
 &\equiv& \int d\omega ~
   \left[ a_0 (\omega) + 2 a_1 (\omega) + 2 a_2 (\omega) \right].
\eea
This partition function includes the many body effect of the
interaction between impurity and the conduction band and is exact
in its form. The approximation comes in when we choose
the approximate Green's functions for the atomic states.

In our simple model, only two symmetry electrons are present: $\Gamma_7$ and
$\Gamma_8$. Their spectral functions (measurable) are defined
as a convolution of two neighboring configurations' spectral functions.
It can be shown that these two interconfiguration atomic spectral
functions are given by
\bea
\rho_{\Gamma_7} (\omega)
 &=& { 1+ e^{-\beta\omega} \over Z_{\rm f} } \int d\zeta ~
   a_0 (\zeta) ~ A_1 (\zeta + \omega) = \rho_{01} (\omega), \\
\rho_{\Gamma_8} (\omega)
 &=& { 1+ e^{-\beta\omega}\over Z_{\rm f} } \int d\zeta ~
   a_1 (\zeta) ~ A_2 (\zeta + \omega) = \rho_{12} (\omega)
\eea
in this conserving approximation\cite{kuramoto}.
{}From now on, we will use the notations
$\rho_{01}, \rho_{12}$ in favor of $\rho_{\Gamma_7}, \rho_{\Gamma_8}$.
This approximation does not include any
vertex correction.

 From the leading bubble diagram\cite{bcw},
the static magnetic susceptibility per Ce$^{3+}$ impurity is
\bea
\chi (T)
 &=& {1\over 3} ~\mu_{\rm eff}^2 ~ \tilde{\chi}_{\rm f} (T),
  ~~ \mu_{\rm eff}^2 = {75 \over 49} ~ \mu_{\rm B}^2, \\
\tilde{\chi} (T)
 &=& - {4 \over Z_{\rm f} } \int d\zeta ~ a_1 (\zeta,T)~
    \hbox{Re} G_1 (\zeta,T).
\eea
Here $\mu_{\rm B}$ is Bohr magneton. The reduced dynamic magnetic
susceptibility is
\bea
\tilde{\chi}^{''} (\omega, T)
 &=& { 1 - e^{-\beta\omega} \over Z_{\rm f} }
    \int d\zeta ~a_1 (\zeta,T) ~ A_1 (\zeta + \omega,T).
\eea
We are going to calculate these physical quantities and transport
coefficients.

\section{Zero temperature analysis.}
 Zero temperature analysis\cite{coxruck,zero} of the NCA integral equations
leads to a qualitative understanding of our model system.
Recently, conformal field theory approach\cite{cft}
calculated the scaling dimensions
exactly for the multi-channel Kondo exchange models with any size of
impurity spin.
In Ref. \cite{coxruck}, the evaluation of the scaling dimensions is extended
to the multichannel Anderson/Coqblin-Schrieffer models using the functional
integral formulation. Both models become congruent when the impurity spin is
$S_I=1/2$.

The main results of the zero temperature analysis of the
NCA integral equations are:
(1) We can find the criterion which predicts whether the ground state is
that of the 1, 2, or 3 channel model.
(2) The Kondo energy scale ($T_0$) can be estimated analytically for the case
$\Gamma_{01} = \Gamma_{12}$. $T_0$ in the one- and two-channel model
parameter regimes is shown to vanish as the $f^2 ~ \Gamma_3$ energy level
approaches that of the $f^0$ configuration.
(3) We obtain the correct scaling dimensions for the overcompensated
cases which agree with the conformal field theory results\cite{ludwig}.
(4) The crossover physics between the parameter regimes for different
numbers of channels can be identified.

 The self consistent NCA integral equations can be transformed into
the differential equations for the flat conduction band
in the wide band limit: $D \gg |\epsilon_{1,2} |$.
We analyze the zero temperature NCA equations in the asymptotic
limit $|\omega - E_0| \ll T_0$. $E_0$ is the threshold energy below
which the pseudo particle Green's functions become purely real.
We introduce the inverse Green's functions and transform the self
energy equations at zero temperature into the coupled non-linear
differential equations.\cite{zero}
\bea
g_0 (\omega)
  &\equiv& - 1/G_0 (\omega); ~~ g_1 (\omega)
   \equiv - 1/G_1 (\omega); ~~ g_2 (\omega)
   \equiv - 1/G_2 (\omega), \\
{d \over d\omega} g_0 (\omega)
  &=& -1 - { 2 \Gamma_{01} \over \pi} { 1 \over g_1 (\omega) };
    ~~ g_0 (-D) = D, \\
{d \over d\omega} g_1 (\omega)
  &=& -1 - { \Gamma_{01} \over \pi} { 1 \over g_0 (\omega) }
    - { 2 \Gamma_{12} \over \pi} { 1 \over g_2 (\omega) };
    ~~ g_1 (-D) = D + \epsilon_1, \\
{d \over d\omega} g_2 (\omega)
  &=& -1 - { 2 \Gamma_{12} \over \pi} { 1 \over g_1 (\omega) };
    ~~ g_2 (-D) = D + \epsilon_2.
\eea
Then $g_0$ and $g_2$ can be shown to be related by
\bea
{g_2 \over \Gamma_{12}}
 &=& { g_0 \over \Gamma_{01} }
   +  \left[ {1\over \Gamma_{01}} - {1 \over \Gamma_{12} } \right] ~
  ( \omega - E_0 ) + \gamma_c, \\
\gamma_c
 &=& {\epsilon_2 - E_0 \over \Gamma_{12}}
  + { E_0 \over \Gamma_{01}}.
\eea
In the zero temperature analysis, it is more convenient to define the
``negative frequency" spectral functions by $a_i (\omega) \equiv
e^{-\beta \omega} ~ A_i (\omega) / Z_{\rm f}$.
These spectral functions vanish above the threshold energy $E_0$ and satisfy
\bea
{d \over d\omega} [~a_0 (\omega) |g_0 (\omega)|^2 ~]
  &=& - { 2 \Gamma_{01} \over \pi} ~ a_1 (\omega), \\
{d \over d\omega} [~a_1 (\omega) |g_1 (\omega)|^2 ~]
  &=& - { \Gamma_{01} \over \pi} ~a_0 (\omega)
    - { 2 \Gamma_{12} \over \pi} ~ a_2 (\omega), \\
{d \over d\omega} [~a_2 (\omega) |g_2 (\omega)|^2~]
  &=& - { 2 \Gamma_{12} \over \pi}~ a_1 (\omega).
\eea
It can be shown from the above relations that
\bea
{d \over d\omega}
  \left[~a_0 (\omega) g_0 (\omega)
  + 2 a_1 (\omega) g_1 (\omega) + 2 a_2 (\omega) g_2 (\omega) ~\right]
  &=& a_0 (\omega) + 2 a_1 (\omega) + 2 a_2 (\omega).
\eea
By integrating this equation from $\omega = -\infty$ to $\omega = E_0$,
we find the additional relation,
\bea
\left[~a_0 (\omega) g_0 (\omega)
  + 2 a_1 (\omega) g_1 (\omega) + 2 a_2 (\omega) g_2 (\omega)
 \right]_{\omega = E_0} &=& 1.
\eea
This identity will be useful in finding the asymptotic behavior of the
``negative frequency" spectral functions. As a corollary, we have another
identity in the asymptotic limit,
\bea
{a_0 (\omega) |g_0 (\omega)|^2 \over \Gamma_{01} }
  &=&  {a_2 (\omega) |g_2 (\omega)|^2 \over \Gamma_{12} }.
\eea
This relation can be proved by using the boundary condition
at $\omega = E_0$.
We calculate the physical atomic spectral functions and the
dynamic magnetic susceptibility defined in section IV.
\bea
\rho_{01} (\omega)
 &=& \int d\epsilon ~ \left[ a_0 (\epsilon) A_1 (\epsilon + \omega)
                        + A_0 (\epsilon) a_1 (\epsilon + \omega) \right], \\
\rho_{12} (\omega)
 &=& \int d\epsilon ~ \left[ a_1 (\epsilon) A_2 (\epsilon + \omega)
                        + A_1 (\epsilon) a_2 (\epsilon + \omega) \right], \\
\tilde{\chi}^{''} (\omega)
 &=& \int d\epsilon ~ \left[ a_1 (\epsilon) A_1 (\epsilon + \omega)
                        - A_1 (\epsilon) a_1 (\epsilon + \omega) \right].
\eea
{}From the $\omega$ dependence of the spectral functions near $\omega = 0$,
we can infer the finite temperature dependence of transport coefficients
as will be discussed below.

 We now discuss the phase diagram in the model parameter space using the
zero temperature analysis. $\gamma_c$ decides the low energy and the low
temperature behaviors of our model Hamiltonian.
$\gamma_c$ measures the relative magnitude of the antiferromagnetic
coupling strengths when the charge fluctuation is removed in the model
Hamiltonian. Noting that $E_0 \approx \epsilon_1 + {\cal O} (V_{01}^2,
V_{12}^2)$, we find
\bea
 \gamma_c &\approx& {2\over \pi N(0)} ~
   \left[ {1\over J_2} - {1\over J_1} \right]
\eea
which illustrates the correspondence to the scaling analysis.
If $\gamma_c$ is greater than zero, the singular behavior
shows up in the $f^0$ Green's function, and not in the $f^2$ Green's
function.
Hence the system will be dominated by the $f^0$ and $f^1$ sector leading to
the one-channel Kondo effect.
When $\gamma_c$ is less than zero, the singular behavior
shows up in the $f^2$ Green's function, and not
in the $f^0$ Green's function.
In this case $f^1$ and $f^2$ sector (the two-channel Kondo physics)
determines the low temperature behavior of
the system.
When $\gamma_c =0$, $f^0$ and $f^2$ become equivalent
asymptotically ($|\omega - E_0| \ll T_0$).
Both Green's functions develop the singular behaviors
at the ground state energy.
In this model parameter space, the three-channel Kondo model
is realized.

The characteristic Kondo energy scale $T_0$ is found from
an integration constant
which connects the low and high energy states. We can obtain the
integration constant  for the case $\Gamma_{12} = \Gamma_{01} (=\Gamma)$.
We will analyze this case in detail and indicate subsequently
how to extend the zero temperature analysis
to the general hybridization strength.

\subsection{The symmetric hybridization limit:
$\Gamma_{12} = \Gamma_{01} (=\Gamma)$. }
 When we take a symmetric hybridization limit
$\Gamma_{12} = \Gamma_{01} (=\Gamma)$, the equations are simplified.
\bea
\gamma_c &=& {\epsilon_2 \over \Gamma}, \\
g_2 &=& g_0 + \epsilon_2, \\
a_0 g_0^2 &=& a_2 g_2^2.
\eea
The last two relations hold true in the asymptotic limit,
$|\omega - E_0| \ll T_0$.
The different ground states then are determined solely by the sign of
the $f^2$ configuration energy relative to the $f^0$ configuration
energy. Removing the variable $\omega$, we can find the
differential equations between the inverse Green's functions.
Integrating from their values at $\omega = -D$ to those at $\omega$, we
find the integration constant
\bea
\exp\left[ {\pi \over 2 \Gamma} ~
    [~ g_0 - g_1 + \epsilon_1 ~] \right]
 &=& \left[ {g_1 \over D} \right] ~
   \left[{g_0 \over D} \right]^{-1/2} ~
   \left[{g_0 + \epsilon_2 \over D} \right]^{-1}
\eea
in the wide conduction band limit, $D >> |\epsilon_{1,2}|$.
We identify three cases for evaluating $T_0$. \\
(1) {\it One-channel case:} $\gamma_c > 0$ or $\epsilon_2 > 0$.
$f^0$ Green's function develops
a singular behavior at the threshold energy, while $f^2$ Green's function
does not. Thus the ground state physics is dominated by the sector
$f^0 - f^1$ leading to the one-channel Kondo effect.
The integration constant and the Kondo temperature are
\bea
{g_1 \over T_0}
 &=& \left[{g_0 \over \Delta} \right]^{1/2} ~
  \left[ 1 + {g_0 \over \epsilon_2} \right] ~
  \exp \left[ {\pi (g_0 - g_1) \over 2 \Gamma} \right], \\
T_0
 &=& D ~ \left[{\epsilon_2 \over D} \right] ~
    \left[{\Gamma \over \pi D}\right]^{1/2} ~
    \exp \left[ {\pi \epsilon_1 \over 2 \Gamma} \right].
\eea
Here $\Delta \equiv \Gamma / \pi$.
We can find the asymptotic behavior of the Green's functions for each
atomic state.
\bea
{g_0 (\omega) \over \Delta}
 &\approx& \tilde{\Omega}^{\alpha_0}, \\
{g_1 (\omega) \over T_0}
  &\approx& \tilde{\Omega}^{\alpha_1}, \\
g_2 (\omega)
 &=& g_0 (\omega) + \epsilon_2, \\
\tilde{\Omega}
 &=& 3{ E_0 - \omega \over T_0 }, \\
\alpha_0 &=& {2 \over 3}; ~~ \alpha_1 = {1 \over 3}.
\eea
Though the scaling dimension is not correct, the estimated Kondo
temperature is correct within order unity and $\chi (0) \sim 1/T_0$.
An interesting observation
is that Kondo temperature vanishes as $\epsilon_2$ tends to zero, i.e.,
the three-channel parameter regime (see the discussion below).

 The detailed derivation of the asymptotic behavior is not included since
NCA does not produce the Fermi liquid ground state in the one-channel
model. We just give a brief summary of the zero temperature analysis
which is relevant to our study. $A_2(\omega)$ vanishes as
$\tilde{\Omega}^{2/3}$ at the threshold  energy, while $A_0$ ($A_1$) is
singular as $\tilde{\Omega}^{-2/3}$ ($\tilde{\Omega}^{-1/3}$)
at $\omega = E_0$.
Thus, the physical spectral function $\rho_{12}$ vanishes
as $|\omega|^{4/3}$ at the Fermi energy, while $\rho_{01}$ is finite.

(2) {\it Two-channel case:} $\gamma_c < 0$ or $\epsilon_2 < 0$.
 In contrast to the one-channel case,
the $f^2$ spectral function has a singular structure leading to two-channel
ground state. The integration constant and the Kondo temperature are
\bea
{g_1 \over T_0}
 &=& \exp\left[{\pi (g_2 - g_1) \over 2 \Gamma}\right] ~
    \left[ 1 + {g_2 \over |\epsilon_2|} \right]^{1/2} ~
     {g_2 \over \Delta}, \\
T_0 &=& D ~ \left[{|\epsilon_2| \over D }\right]^{1/2} ~
   \left[ {\Gamma \over \pi D } \right] ~
   \exp \left[{\pi (\epsilon_1 - \epsilon_2)
      \over 2 \Gamma } \right].
\eea
Note that the Kondo temperature vanishes with $\epsilon_2 \to 0$.
We can find the asymptotic behavior following
the standard zero temperature analysis.
\bea
g_0 (\omega) &=& g_2 (\omega) + |\epsilon_2|, \\
{g_1 (\omega) \over T_0}
 &=& \tilde{\Omega}^{1/2} ~[~ 1 + c_1 \tilde{\Omega}^{1/2}
       + {\cal O} (\tilde{\Omega})~], \\
{g_2 (\omega) \over \Delta}
 &=& \tilde{\Omega}^{1/2} ~[~ 1 + c_2 \tilde{\Omega}^{1/2}
       + {\cal O} (\tilde{\Omega})~], \\
\tilde{\Omega} &=& 4 {E_0 -\omega \over T_0}, \\
c_1 &=& {1\over 6} ~ \left[ 2\left(1 + {\Delta \over |\epsilon_2|}\right)
               - {T_0 \over \Delta} \right]; ~~
 c_2 = -{1\over 6} ~ \left[ \left(1 + {\Delta \over |\epsilon_2|}\right)
               - 2 {T_0 \over \Delta} \right].
\eea
The asymptotic behavior above the ground state energy $E_0$ can be obtained
from the expressions below $E_0$ by the analytic continuation.
Furthermore, we find for the ``negative frequency"  spectra that
\bea
{d \over d\tilde{\Omega} } a_0 g_0^2
 &=& {T_0 \Delta \over 2} ~ a_1, \\
{d \over d\tilde{\Omega} } a_1 g_1^2
 &=& {T_0 \Delta \over 2} ~ [~ {1\over 2} a_0 + a_2~], \\
{d \over d\tilde{\Omega} } a_2 g_2^2
 &=& {T_0 \Delta \over 2} ~ a_1,
\eea
which implies for $|\omega - E_0| \ll T_0$
\bea
a_0 &=& {\Delta \over 4 |\epsilon_2|^2} ~\tilde{\Omega}^{1/2}
      ~[~ 1 - x_0 \tilde{\Omega}^{1/2} + {\cal O} (\tilde{\Omega})~], \\
a_1 &=& {1\over 4 T_0} ~\tilde{\Omega}^{-1/2}
      ~[~ 1 - x_1 \tilde{\Omega}^{1/2} + {\cal O} (\tilde{\Omega})~], \\
a_2 &=& {1\over 4 \Delta} ~\tilde{\Omega}^{-1/2}
      ~[~ 1 - x_2 \tilde{\Omega}^{1/2} + {\cal O} (\tilde{\Omega})~], \\
x_0 &=& {2\over 3} (2c_1 + c_2) + 2 {\Delta \over |\epsilon_2|}
   = {1\over 3} ~\left(1 + {7\Delta \over |\epsilon_2|} \right), \\
x_1 &=& {4\over 3} (2c_1 + c_2) =
  {2\over 3} ~\left(1 + {\Delta \over |\epsilon_2|} \right), \\
x_2 &=& {4\over 3} (c_1 + 2c_2) = {2\over 3} ~ {T_0 \over \Delta}.
\eea
Note that $x_{0,1,2} > 0$.
Now we can find the asymptotic behavior of the pseudo particle
spectral functions.
\bea
A_0 (\omega)
 &=& {\Delta \over \pi |\epsilon_2|^2} ~ \theta (\omega - E_0) ~
  \left[ |\tilde{\Omega}|^{1/2} + {\cal O} (|\tilde{\Omega}|^{3/2} )
  \right], \\
A_1 (\omega)
 &=& {1\over \pi T_0} ~ \theta (\omega - E_0) ~
  \left[ |\tilde{\Omega}|^{-1/2} + {\cal O} (|\tilde{\Omega}|^{1/2} )
  \right], \\
A_2 (\omega)
 &=& {1\over \pi \Delta} ~ \theta (\omega - E_0) ~
  \left[ |\tilde{\Omega}|^{-1/2} + {\cal O} (|\tilde{\Omega}|^{1/2} )
  \right].
\eea
As expected, $A_0 (\omega)$ vanishes at the threshold energy and does not
develop any singular behavior. On the other hand, $A_1 (\omega)$ and
$A_2 (\omega)$ are singular at the threshold energy.
Finally the physical spectral functions are given by in the asymptotic
limit.
\bea
\rho_{01} (\omega > 0)
 &=& {\Delta \over 32 |\epsilon_2|^2} ~
  \left[\tilde{\omega}
    - {8x_0 \over 3\pi} ~ \tilde{\omega}^{3/2} + \cdots \right], \\
\rho_{01} (\omega < 0)
 &=& {\Delta \over 32 |\epsilon_2|^2} ~
  \left[|\tilde{\omega}|
    - {4x_1 \over 3\pi} ~ |\tilde{\omega}|^{3/2} + \cdots \right], \\
\rho_{12} (\omega > 0)
 &=& {1\over 16 \Delta} ~
  \left[ 1 - {2x_1 \over \pi} ~ \sqrt{\tilde{\omega}} + \cdots \right], \\
\rho_{12} (\omega < 0)
 &=& {1\over 16 \Delta} ~
  \left[ 1 - {2x_2 \over \pi} ~ \sqrt{|\tilde{\omega}|} + \cdots \right], \\
\chi^{''} (\omega)
 &=& {\mbox{sign} (\omega) \over 16 \Delta} ~
  \left[ 1 - {2x_1 \over \pi} ~ \sqrt{|\tilde{\omega}|} + \cdots \right], \\
\tilde{\omega} &=& 4 {\omega \over T_0}.
\eea
$\rho_{01} (\omega)$ vanishes at $\omega = 0$ and increases as $|\omega|$
near the fermi level. This spectral depletion at the fermi level
is also confirmed in the finite temperature NCA calculation.
$\rho_{12}$ is peaked right at the fermi level and has more spectral weight
below than above the fermi level since $x_2 < x_1$ for $T_0 < \Delta$.
The finite temperature NCA results also confirm this observation.
The spectral functions become non-analytic at the Fermi level at zero
temperature.
Note that the dynamic magnetic susceptibility is step function like at
$\omega = 0$, which is none other than marginal fermi liquid behavior
\cite{coxruck}.
{}From the asymptotic form of the zero temperature spectral functions, we
can infer the finite temperature dependence of the resistivity using the
Kubo formula
\bea
{1\over \rho (T)}
 &=& {ne^2 \over m} ~ \int d\omega \tau (\omega) ~
   \left(-{\partial f \over \partial \omega} \right).
\eea
For $\omega, T \ll T_{0}$, the relaxation rate for the conduction electron
is approximately
\bea
{1\over \tau (\omega)}
 &\propto& \rho_{12} (\omega, T).
\eea
Near the zero temperature, we may replace $\rho_{12} (\omega, T)$ by our zero
temperature one and find the $\sqrt{T}$ temperature dependence.
In fact one complication arises due to the angular averaging.
Since the angular harmonics conjugate to $\rho_{12}$ is positive definite,
still the above simple argument applies. In the one-channel case, we cannot
use this kind of simple argument to find the low temperature behavior of
resistivity.
The scaling dimensions agree with those from the conformal field theory
approach.

 We summarize the low temperature properties:
(i) The atomic spectral function $\rho_{12}$ is peaked right at the Fermi
level independent of the occupancy, while $\rho_{01}$ vanishes at
$\omega =0$ and is depleted near the fermi level;
(ii) The dynamic magnetic susceptibility is step function-like
at zero frequency;
(iii) The resistivity obeys the scaling behavior
$\rho (T) = \rho (0)~[~ 1 - a \sqrt{T/T_0} ~]$
near zero temperature.
 All these results will be shown consistent with numerical NCA calculations
at finite temperatures.

(3) {\it Three-channel case:} $\gamma_c = 0$ or $\epsilon_2 = 0$.
 In this case, the $f^0$ and
$f^2$ configurations are equivalent asymptotically leading to three-channel
ground state. The integration constant and the Kondo temperature are
\bea
{g_1 \over T_0}
 &=& \left[{g_0 \over \Delta} \right]^{3/2} ~
    \exp \left[ {\pi \over 2 \Gamma} ~ (g_0 - g_1) \right], \\
T_0 &\approx& D ~
  \left[{\Gamma \over \pi D }\right]^{3/2} ~
   ~ \exp \left[{\pi \epsilon_1 \over 2 \Gamma } \right].
\eea
See the Appendix for the detailed derivation of asymptotic behavior.
This model case is not different from the analysis of the general
overcompensated models. We simply list the asymptotic behavior here.
\bea
g_0 (\omega) &=& g_2 (\omega), \\
{g_1 (\omega) \over T_0}
 &=& \tilde{\Omega}^{3/5} ~ \left[
    1 + c_1 ~ \tilde{\Omega}^{2/5} + d_1 ~ \tilde{\Omega}^{3/5}
     + \cdots ~\right], \\
{g_2 (\omega) \over \Delta}
 &=& \tilde{\Omega}^{2/5} ~ \left[
    1 + c_2 \tilde{\Omega}^{2/5} + d_2 ~ \tilde{\Omega}^{3/5}
    + \cdots ~\right], \\
\tilde{\Omega}
 &=& 5 { E_0 - \omega \over T_0 }, \\
c_1 &=& {2\over 7}, ~~ d_1 = - {T_0 \over 8\Delta }; ~~
 c_2 = - {1\over 7 }, ~~ d_2 = {T_0 \over 4\Delta }.
\eea
The nagative spectral functions are
\bea
a_1 (\omega)
 &=& {1 \over 5T_0}~ \tilde{\Omega}^{-3/5} ~
  \left[1 - x_1 ~ \tilde{\Omega}^{2/5} - y_1 ~ \tilde{\Omega}^{3/5}
        + \cdots \right] \theta (\tilde{\Omega}), \\
a_2 (\omega)
 &=& {1 \over 5\Delta}~ \tilde{\Omega}^{-2/5} ~
  \left[1 - x_2 ~ \tilde{\Omega}^{2/5} - y_2 ~ \tilde{\Omega}^{3/5}
        + \cdots \right]
  \theta (\tilde{\Omega}), \\
x_1 &=& 2c_1 = {4\over 7}, ~~y_1 = 0; ~~x_2 = 0, ~~
 y_2 = 2 d_2 = {T_0 \over 2\Delta}.
\eea
And the pseudo particle spectral functions are
\bea
A_1 (\omega)
 &=& {\sin (3\pi/5) \over \pi T_0}~
     |\tilde{\Omega}|^{-3/5} ~
  \left[ 1 - x_1 ~\cos (2 \pi/5) ~
       |\tilde{\Omega}|^{2/5} + \cdots \right]
    \theta (-\tilde{\Omega}), \\
A_2 (\omega)
 &=& {\sin (2 \pi/5) \over \pi \Delta}~
     |\tilde{\Omega}|^{-2/5} ~
  \left[ 1 - y_2 ~\cos (2\pi/5) ~
       |\tilde{\Omega}|^{3/5} + \cdots \right]
    \theta (-\tilde{\Omega}).
\eea
{}From the pseudo particle spectral functions, we can find the
physical spectral functions in the asymptotic limit.
\bea
\rho_{01} (\omega)
 &=& \rho_{12} (-\omega), \\
\rho_{12} (\omega > 0)
 &=& {\sin(2\pi/5) \over 25 \pi \Delta}
  ~[~ B(2/5,3/5)
   - x_1 ~ B(3/5,4/5)~ |\tilde{\omega}|^{2/5} \nonumber\\
 && \hspace{2.0cm} - y_2~ \cos(2\pi/5)~ B(2/5,6/5) ~|\tilde{\omega}|^{3/5}
   + \cdots ~], \\
\rho_{12} (\omega < 0)
 &=& {\sin(3\pi/5) \over 25 \pi \Delta}
  ~[~ B(2/5,3/5)
   - x_1~ \cos(2\pi/5)~ B(3/5,4/5) ~|\tilde{\omega}|^{2/5} \nonumber\\
 && \hspace{2.0cm}- y_2 ~ B(2/5,6/5) ~ |\tilde{\omega}|^{3/5} + \cdots ~], \\
\tilde{\chi}^{''} (\omega)
 &=& \mbox{sign} (\omega) ~ {\sin(3\pi/5) \over 25 \pi T_0}
  ~[~ B(2/5,2/5) ~ |\tilde{\omega}|^{-1/5}  \nonumber\\
 && \hspace{2.0cm}- x_1 ~[~1 + \cos(2\pi/5)~]~ B(2/5,4/5)
    ~ |\tilde{\omega}|^{1/5} + \cdots ~], \\
\tilde{\omega}
 &=& 5 ~ {\omega \over T_0}.
\eea
Here $B(p,q)$ is the Beta function.
The spectral functions become non-analytic at the Fermi level at zero
temperature.
One important observation is that the power laws agree with those
obtained from the conformal field theory treatments for the overcompensated
multichannel $S=1/2$ models. Since $\rho_{01}$ and $\rho_{12}$ are
equivalent in the asymptotic limit, the angular functions are factored
out in the conduction electron scattering time. Thus we can read off the
power law of the resistivity, $\alpha = 2/5$.

Again, the low temperature properties are summarized below.
(i) The atomic spectral function is peaked right at the Fermi level
independent of the occupancy;
(ii) The dynamic magnetic susceptibility diverges as $\omega^{-1/5}$
at zero frequency;
(iii) The resistivity obeys the scaling behavior
$\rho (T) = \rho (0)~[~ 1 - a (T/T_0)^{0.4} ~]$
near zero temperature. All these results will be shown consistent
with numerical NCA calculations at finite $T$.

\subsection{Crossover physics.}
 We expect the smooth crossover from the high temperature
regime to the low temperature regime where the relevant channel number
physics shows up. We can estimate the crossover temperature in the one- and
two-channel model parameter regime using the zero temperature
analysis. Noting that the asymptotic analysis assumes $g_0 << \epsilon_2$
for the one-channel regime and $g_2 << |\epsilon_2|$ for the two-channel
regime (see the integration constant), we can estimate the crossover energy
scale below which one- or
two-channel physics is realized. In the one-channel case, the integration
constant can be rewritten as
\bea
{g_1 \over T_0^{(3)} }
 &=& \left[ 1 + {\epsilon_2 \over g_0} \right] ~
  \left[{g_0 \over \Delta} \right]^{3/2} ~
  \exp \left[ {\pi (g_0 - g_1) \over 2 \Gamma} \right].
\eea
We can see that the relative magnitude of $g_0$ and $\epsilon_2$
determines the differing channel behavior. The crossover
energy scale can be defined by the relation $g_0 = \epsilon_2$.
The crossover temperature is found in the one-channel case
\bea
T_{\rm x}^{(1)}
 &=& {T_0 \over 3} ~ \left[{\epsilon_2 \over \Delta} \right]^{3/2}.
\eea
And in the two-channel case, we find the crossover temperature
\bea
T_{\rm x}^{(2)}
 &=& {T_0 \over 4} ~ \left[{|\epsilon_2| \over \Delta} \right]^2.
\eea
Substituting the Kondo energy scale, we can see that the crossover
energy scales are
\bea
T_{\rm x}^{(1)}
 &=& {1\over 3} ~ \left[ \epsilon_2 \over \Delta \right]^{5/2} ~
     D ~ \left[{\Gamma \over \pi D} \right]^{3/2} ~
     \exp \left[{\pi \epsilon_1 \over 2\Gamma} \right], \\
T_{\rm x}^{(2)}
 &=& {1\over 4} ~ \left[ |\epsilon_2| \over \Delta \right]^{5/2} ~
     D ~ \left[{\Gamma \over \pi D} \right]^{3/2} ~
     \exp \left[{\pi (\epsilon_1 - \epsilon_2) \over 2\Gamma} \right].
\eea
Note that the crossover temperature vanishes as $|\epsilon_2|^{5/2}$
with $\epsilon_2 \to 0$. We intentionally write the crossover temperature
as the prefactor times the corresponding three-channel Kondo temperature.
If this prefactor is greater than 1, the system will not display the
three-channel behavior with decreasing temperatures but will flow directly
to the one- or two-channel fixed point.
When this prefactor is less than 1, the system
will display the three-channel behavior before finally flowing to each
channel fixed point.
In the one-channel regime, the system flows directly from the
high temperature regime (local moment regime) to the one-channel fixed
point when $\epsilon_2 >> \Delta$ with decreasing temperatures. In the
opposite limit, $\epsilon_2 << \Delta$, the system will flow from the
local moment regime to the three-channel regime, and finally to
the one-channel fixed point after passing through the crossover temperature.
In the two-channel regime the same argument can be applied as in
the one-channel case.

\subsection{General asymmetric hybridization case.}
 Though the Kondo energy scale can not be estimated analytically
for the general case $\Gamma_{01} \neq \Gamma_{12}$,
still the asymptotic behavior ($\omega \leq E_0$) can be sorted out. Since
the physics is the same as the above analysis,
we shall just write the asymptotic
properties of Green's functions. \\
(1) $\gamma_c > 0$:
\bea
g_0 (\omega)
 &\approx& {\Gamma_{01} \over \pi}
  \left[ 3{ E_0 - \omega \over T_0 } \right]^{\alpha_0},
  \\
g_1 (\omega)
  &\approx& T_0 \left[ 3
   { E_0 - \omega \over T_0 } \right]^{\alpha_1}, \\
g_2 (\omega)
  &=& {\Gamma_{12} \over \Gamma_{01}} ~ g_0 (\omega)
  +  \left[ {\Gamma_{12} \over \Gamma_{01}} - 1 \right] ~
  ( \omega - E_0 ) + \gamma_c ~ \Gamma_{12} , \\
\alpha_0 &=& {2 \over 3}; ~~
  \alpha_1 = {1 \over 3}.
\eea
\\
(2) $\gamma_c < 0$:
\bea
g_0 (\omega)
  &=& {\Gamma_{01} \over \Gamma_{12}} ~ g_2 (\omega)
  +  \left[ {\Gamma_{01} \over \Gamma_{12}} - 1 \right] ~
  ( \omega - E_0 ) + |\gamma_c|~\Gamma_{01}, \\
g_1 (\omega)
  &\approx& T_0 \left[ 4
   { E_0 - \omega \over T_0 } \right]^{\alpha_1}, \\
g_2 (\omega)
 &\approx& {\Gamma_{12} \over \pi}
  \left[ 4{ E_0 - \omega \over T_0 } \right]^{\alpha_2},
  \\
\alpha_1 &=& {1 \over 2}; ~~
  \alpha_2 = {1 \over 2}.
\eea
\\
(3) $\gamma_c = 0$: In this case, $f^0$ and $f^2$ are equivalent
asymptotically.
\bea
g_2 (\omega)
 &=& {\Gamma_{12} \over \Gamma_{01}} ~ g_0 (\omega)
   + \left[ {\Gamma_{12} \over \Gamma_{01}} - 1 \right] ~ (\omega - E_0 )
   \\
{d \over d\omega} g_0 (\omega)
 &=& - 1 - { 2 \Gamma_{01} \over \pi} { 1 \over g_1 (\omega) };
    ~~ g_0 (-D) = D, \\
{d \over d\omega} g_1 (\omega)
 &=& -1 - {\Gamma_{01} \over \pi}
   ~\left[ { 1 \over g_0 (\omega) }
     + { 2 \over g_0 (\omega) + (1- \Gamma_{01}/ \Gamma_{12} )
       (\omega - E_0) }
   \right]; ~~ g_1 (-D) = D + \epsilon_1.
\eea
Then the asymptotic form follows
\bea
g_0 (\omega)
 &\approx& {\Gamma_{01} \over \pi}
  \left[ 5{ E_0 - \omega \over T_0 } \right]^{\alpha_0},
  \\
g_1 (\omega)
 &\approx& T_0 ~ \left[ 5{ E_0 - \omega \over T_0 } \right]^{\alpha_1}, \\
\alpha_0 &=& {2 \over 5}; ~~
  \alpha_1 = {3 \over 5}.
\eea
Though the Kondo energy scale cannot be evaluated analytically, it can
be estimated from the numerical
NCA calculation of, e.g., magnetic susceptibility
and fitting it onto the universal results.

\section{Numerical Analysis.}
 In this section we present results from our full numerical study at
finite temperatures.
 We studied the model for the parameters listed in Table \ref{modelpar}.
This covers the
one-, two-, and three-channel Kondo regimes. For simplicity, we have chosen
the same hybridization strength in the $f^0 - f^1$ and the $f^1 - f^2$
sectors. The relevant channel-number-dependent physics can then be studied by
varying the
relative position of the $f^0$ and the $f^2$ configuration energies.
This simple choice of the hybridization further makes it possible to find
the characteristic Kondo energy scales analytically, which is described
in the previous section in detail.

 Our main results are:
(1) The magnetic susceptibility shows a scaling behavior and
agrees well with the exact Bethe ansatz results in the two-
and three-channel parameter regimes;
(2) The NCA results of residual entropy in the two-, and three-channel
models agree with the exact ones within order ${\cal O} (1/N)$. The Kondo
anomaly peak in the specific heat also agrees with the exact one
in its magnitude;
(3) The thermopower is a diagnostic to display the ground states
for different numbers of relevant channels for our model;
(4) The dynamic magnetic susceptibility is distinct between the one-channel
model and overcompensated (two- and three-channel) models;
(5) Due to the simplifying features of our model, the resistivity shows
a bendover at low temperatures in the one-channel parameter regime.
The resistivity in the two- and
three-channel parameter regimes shows temperature dependences
near $T = 0K$ agreeing with the conformal field theory results;
(6) We confirmed in detail that NCA is a valid
numerical self-consistent non-perturbative method
in studying the overcompensated
multi-channel $S_I=1/2$ Anderson model at $T>0$.

\subsection{Entropy and specific heat.}
 The entropy and the specific heat due to the magnetic impurity can be
calculated from the free energy obtained in NCA through numerical
differentiation.
These thermodynamic quantities include very important information
about the nature of the ground state. We can estimate the characteristic
energy scales in the Kondo models from the temperature variation of
the entropy. In general, the entropy will increase with increasing
temperatures, until the frozen impurity degrees of freedom are released.
Our model Hamiltonian is expected to have the entropy $S = k_B \ln 5$
at high enough temperatures, while the residual entropy at zero temperature
will show behavior expected for different numbers of relevant channels.

 The entropy and the specific heat  are displayed in Fig.~\ref{entropy} and
\ref{spheat} for the model
parameter sets studied here. In the one-channel case, the Kondo anomaly
peak is well separated from the Schottky anomaly peak coming from the
interconfiguration excitations. The Kondo anomaly peak has a magnitude
comparing well with the exact results\cite{sacramento}
for the Kondo exchange model.
No residual entropy remains with $T\to 0$.
In the two-channel case, the Kondo anomaly peak is not clearly
separated from the charge fluctuation peak for most of our
model parameters.
The residual entropy agrees within 5 \% with the exact
one\cite{sacramento}
and the discrepancy can be explained by the ${\cal O}(1/N^2)$ and
higher order corrections. To see this, we note that the entropy
for spin $1/2$ conduction electrons has an explicit dependence
on the impurity spin degeneracy\cite{cft,ludwig}.
Strictly speaking, the NCA results are valid for $N$-fold impurity
spin and $M$-fold channel degeneracies as $N\to \infty$ with
$N/M$ fixed. Hence it is natural to expect ${\cal O} (1/N^2)$
corrections to the entropy through the neglected vertex
corrections.
In the three-channel case, the Kondo anomaly peak is reduced
further due to the increased residual entropy and is almost merged with
the charge fluctuation contribution for model set 4. For model set 5,
we can see a weak indication of separation. The residual entropy is increased
compared to that of two-channel case and its magnitude is again a little bit
smaller than the exact one\cite{sacramento}
due to the neglect of the higher
order contribution in $1/N$ expansion.

\subsection{Static magnetic susceptibility.}
 The static magnetic susceptibility is a direct indicator of
the nature of the ground state for the magnetic Kondo model.
As is well documented,
the magnetic susceptibility diverges logarithmically
$\chi(T) \propto \log (T_0 /T)$ for the two-channel $S_I=1/2$
magnetic Kondo model at zero temperature,
and diverges algebraically for the three
or the higher channel $S_I=1/2$ Kondo model
($\chi (T) \propto \left[ T_0 /T \right]^{1-2\Delta_n};
{}~~\Delta_n = 2/(n+2)$: $n$ is the channel number ($\geq 3$))
\cite{cft,ludwig,sacramento}.
The one-channel Kondo model has
a constant magnetic susceptibility at zero temperature ($\chi(0) \sim
1/T_0$). From the NCA, using the leading order bubble diagram\cite{bcw},
the magnetic susceptibility for one Ce$^{3+}$ impurity is given by
\bea
\chi (T)
 &=& {1\over 3} \mu_{\rm eff}^2 ~ \tilde{\chi}  (T),
  ~~ \mu_{\rm eff}^2 = {75 \over 49} ~ \mu_{\rm B}^2, \\
\tilde{\chi} (T)
 &=& - {4 \over Z_{\rm f} } \int d\zeta ~ a_1 (\zeta,T)~
    \hbox{Re} G_1 (\zeta,T).
\eea
Here $\mu_{\rm B}$ is the Bohr magneton.

 Our numerical results for
$\tilde{\chi}  (T)$ clearly show the right tendency
for each possible low $T$ channel number. The magnetic susceptibility in the
one-channel
regime, for the model parameter sets 6, 7, 8 shows
 approximate scaling behavior and clearly has the negative
curvature at low temperatures indicating approach to the Fermi liquid ground
state. The deviation of the scaling behavior
at low temperatures seems to come from the pathological behavior of NCA
in this one-channel case\cite{zero}.

The $\tilde{\chi} (T)$ curves in the two-channel  regime, for
the model parameter sets 1, 2, 3, also show scaling
behavior and diverge logarithmically at low temperatures (Fig.~\ref{chi}).
Our results
are compared with those of Ref.\cite{sacramento}. The fitting to
the exact Bethe ansatz numerical results\cite{sacramento} is quite good.
We believe that the high temperature deviation comes
from the $M=3$ to $M=2$ crossover physics described in the previous section.
Note that Bethe ansatz results are for the pure two-channel  s-d
exchange model. To get the fitting to the Bethe ansatz, we slide
the temperature axis to find $T_K = 0.3 \times T_0$
($T_K$ from the Bethe Ansatz). Here $T_0$ is the
Kondo energy scale estimated from the zero temperature analysis.

At high temperatures we cannot distinguish
the different numbers of relevant channels clearly.
Note that distinct physics for different
numbers of relevant channels
shows up at low temperatures below the crossover temperature which
was estimated in the previous section. This observation is supported
by results from the three-channel  model parameter sets. The high temperature
deviation from scaling is very weak in this case which will not show
the crossover physics.
The magnetic susceptibility in the three-channel  model also shows
a scaling behavior.
Since the three-channel  case lies exactly on the boundary between
the one-channel  and the two-channel  regime, we probed the three-channel
case by varying the position of the bare $f^1$ configuration energy while
the two excited configurations $f^0, f^2$ energies are kept equal.
Fitting to the exact results is quite good. The Kondo energy scale $T_0$
estimated from the zero temperature analysis agrees with that in the
exact Bethe ansatz ($T_K$ of Ref\cite{sacramento}).

\subsection{Spectral functions.}
 The interconfiguration spectral functions show
a distinct behavior for different numbers of relevant channels
near the the Fermi level ($\omega=0$) depending on the model
parameters. In our simple model, we have two symmetry
atomic spectral functions of $\Gamma_7$ and $\Gamma_8$:
$\rho_{\Gamma_7} (\omega) = \rho_{01} (\omega)$
and $\rho_{\Gamma_8} (\omega) = \rho_{12} (\omega)$.
\bea
\rho_{01} (\omega)
 &=& { 1+ e^{-\beta\omega} \over Z_{\rm f} } \int d\zeta ~
   e^{-\beta \zeta} ~ A_0 (\zeta) ~ A_1 (\zeta + \omega) \\
\rho_{12} (\omega)
 &=& { 1+ e^{-\beta\omega}\over Z_{\rm f} } \int d\zeta ~
   e^{-\beta \zeta} ~ A_1 (\zeta) ~ A_2 (\zeta + \omega)
\eea

In the one-channel model parameter regime (see Fig.~\ref{rho1ch}),
 $\rho_{01} (\omega)$ develops
the Kondo resonance peak just above the Fermi level and
$\rho_{12} (\omega)$ is depleted near $\omega =0$ and tends to zero
with $T\to 0$ at $\omega = 0$. This confirms our zero temperature
analysis in section IV. For comparison, note that the Kondo
resonance amplitude is big compared to the two-channel or three-channel
cases.

In the two-channel regime (see Fig.~\ref{rho2ch}),
$\rho_{12} (\omega)$ is peaked below $\omega =0$  and its peak
position tends to $\omega = 0$ with decreasing temperature. Note that
the Kondo resonance amplitude is reduced compared to the one-channel
Kondo resonance amplitude. On the other hand,
$\rho_{01}(\omega)$ is depleted near $\omega = 0$.

In the three-channel parameter regime (see Fig.~\ref{rho3ch}),
the two spectral functions are equivalent asymptotically when
$\omega \to -\omega$ transformation is accounted for. Peak positions
tend to zero with decreasing temperature. Both develop
the Kondo resonance peak with further reduced amplitude.

 As mentioned above, the positions of the Kondo resonance peak
(see Fig. \ref{kondopeak})
show a distinct behavior for different numbers of relevant channels.
The peak position saturates to a constant value in the one-channel
cases while it vanishes in the two- and three-channel cases with
decreasing temperature. In addition, the detailed functional
forms of the Kondo resonance peak
are different depending on the relevant channel numbers.
In the overscreened cases ($M=2,3$), the peak structure becomes
non-analytic with decreasing temperature as shown in the
zero temperature analysis of the NCA integral equations.
Finite temperature washes out this non-analytic behavior
at the Kondo resonance peak. In the one-channel case,
the atomic spectral functions remain analytic down to
zero temperature.

 Since we are not considering all the atomic energy levels, the
full atomic spectral functions (measured from the photoemission
experiments) cannot be defined in our simple model. Though
the high energy physics of real systems can not be properly
treated within our simple model Hamiltonian, the low energy or
low temperature properties can be studied
using the restricted spectral functions.
Note that low temperature/energy physics is governed by the
Kondo resonance peak development.
Though the spectral depletion is found in the interconfiguration
spectral functions, we do not expect that the photoemission spectroscopy
can observe this feature unless it can distinguish the atomic electron
symmetry. Measurable atomic spectral functions
are shown in Fig.~\ref{rhof} for each relevant channel case.

\subsection{Dynamic magnetic susceptibility.}
 The dynamic magnetic susceptibility
measures the magnetic excitation structure.
Since the properties of the magnetic excitations are related to the
interaction of the local magnetic moment with the conduction electrons,
the relevant channel number will determine the nature of the magnetic
excitations.
The dynamic magnetic susceptibility is expected to be distinct for
each number of relevant channels.
We have already seen this channel dependence
in the static magnetic susceptibility.

 Dynamic magnetic susceptibility is defined as the spin-spin correlation
function and can be measured directly from the neutron scattering
experiments. From the leading bubble diagram, the reduced dynamic magnetic
susceptibility is
\bea
\tilde{\chi}^{''} (\omega, T)
 &=& { 1 - e^{-\beta\omega} \over Z_{\rm f} }
    \int d\zeta ~a_1 (\zeta,T) ~ A_1 (\zeta + \omega,T).
\eea
Our reduced static magnetic susceptibility is related to the above magnetic
response function by the Kramer-Kronig relation:
\bea
\tilde{\chi} (T)
 &=& 2 \int {d\omega \over \pi} ~
     { \tilde{\chi}^{''} (\omega, T) \over \omega }.
\eea
The neutron scattering experiments measure the structure function
$S(\omega, T) \propto ~[~b(\omega) + 1~]~\chi^{''} (\omega, T)$.
The dynamic susceptibility can be quantitatively characterized
by its `linewidth' dependence on temperature in addition to its
overall functional shape.
We may define the linewidth $\Gamma (T)$
by the peak position of $\chi^{''} (\omega, T)$, which
can be measured directly in the inelastic neutron
scattering experiments.

The variation of $\chi^{''} (\omega, T)$ and $\Gamma (T)$
with temperature are displayed in Fig.~\ref{dchi} and \ref{gamma}.
For the dynamic magnetic susceptibility, we display the reduced
form $\tilde{\chi}^{''} (\omega,T) / \tilde{\chi}^{''} (\Gamma(T),T)$
as a function of $\omega/\Gamma (T)$.
The distinct behavior for different number of relevant channels
is clearly evidenced: in the one-channel
regime, the $\tilde{\chi}^{''} (\omega, T)$ curves converge with decreasing
temperatures (not shown). This is clearly supported by the saturation of
$\Gamma (T)$ at low temperatures. In the two- and three-channel regimes,
$\Gamma (T)$ vanishes algebraically (close to linear)
and the dynamic magnetic susceptibility does not
converge (not shown) in contrast to the one-channel case,
but instead develops non-analytic behavior at $\omega =0$.
The reduced dynamic magnetic susceptibilities (defined above,
see Fig.~\ref{dchi})
roughly show a scaling behavior between two extrema.

The physics of the linewidth is quite important in understanding the
nature of the magnetic spin screening. The distinct behavior for different
numbers of channels
shows up at low temperatures below $T_0$. The impurity
spin flipping time ($\tau_{\rm f}$) due to the hybridization will be given
by the inverse of the linewidth, $\tau_{\rm f} \sim 1/\Gamma (T)$.
On the other hand,
thermally excited conduction electrons close to the impurity site
will pass through it in a time $\tau_c$ of
order of $1/T$ from the uncertainty principle.
At high temperatures above the Kondo temperature, $\tau_{\rm f} \gg \tau_c$.
Thus the impurity spin rarely flips while the conduction electrons
pass by the impurity site. Hence the Curie behavior is expected
in the magnetic susceptibility. At low temperatures below the Kondo
temperature, differing channel number physics shows up. In the one-channel
regime, $\tau_{\rm f} \ll \tau_c$ and the impurity spin flips
vigorously averaging out its spin to zero
and leading to the Pauli behavior. In the two- and three-channel regimes,
$\tau_{\rm f} \approx \tau_c$ and spin screening is
not complete
leading to the non-Fermi liquid ground state. This interpretation
agrees with the low temperature behaviors of magnetic susceptibility.

\subsection{Transport coefficients.}
Using the Kubo formula\cite{bcw} in a dilute impurity limit, where
the inter-impurity correlation can be neglected, we have calculated
the transport coefficients, resistivity and thermopower.
The anisotropic conduction electron scattering rate is
\bea
\tau^{-1} (\hat{k}, \omega)
 &=& {8\pi n_{\rm imp} \over N(0) }
  ~[~\Gamma_{01}~\Theta_7^{(5/2)} (\hat{k}) ~ \rho_{01} (\omega)
    + \Gamma_{12}~\Theta_8^{(5/2)} (\hat{k}) ~ \rho_{12} (\omega) ~],
   \\
\Theta_7^{(5/2)} (\hat{k})
 &=& {1\over 16\pi} ~[~ 6 - \Phi (\hat{k}) ~]; ~~
    \Theta_8^{(5/2)} (\hat{k}) = {1\over 16\pi}
     ~[~ 6 + \Phi (\hat{k}) ~], \\
\Phi (\hat{k})
 &=& 15 \cos^4 \theta - 10 \cos^2 \theta + 1
    + 5\sin^4 \theta \cos^2 2\varphi.
\eea
Here $n_{\rm imp}$ is the impurity concentration.
Note that the conduction electron scattering rate contains the
hybridization since the interconfiguration spectral functions are
involved. The crystal harmonics are normalized such that the integration of
$\Theta_7^{(5/2)} (\hat{k})$ and $\Theta_8^{(5/2)} (\hat{k})$
over the solid angle leads to 1 and 2, respectively.
Note that the the crystal harmonic $\Theta_7^{(5/2)} (\hat{k})$
vanishes at ``hot spot" angles of
$(\theta, \varphi) = (0,-)$, $(\pi, -)$, $(\pi/2, 0)$,
$(\pi/2, \pi/2)$,$(\pi/2, \pi)$,$(\pi/2, 3\pi/2)$, while $\Theta_8^{(5/2)}
(\hat{k})$ is positive definite.
This feature, combined with the dip structure of
the interconfiguration spectral function
($f^1 - f^2$ or $\rho_{\Gamma_8} (\omega)$),
leads to a bendover in resistivity at low temperatures in the one-channel
parameter regime within our simplified model. We believe this feature
will go away in more realistic models.
For example, another contribution to the
$\Gamma_8$ atomic spectral function comes from the convolution between
$f^0$ and $f^1~J=5/2~\Gamma_8$. This spectral function will not be
depleted at the fermi level, but instead will build up its spectral
weight due to the weak Kondo resonance structure just above $\omega =0$.
Thus the low temperature bendover in the resistivity
will disappear.

 The channel-dependence of the spectral functions, combined with the
angular dependence, leads to a distinct behavior in the angle-averaged
conduction electron lifetime for different numbers of channels.
The spectral depletion becomes pronounced
in the one-channel regime due to the angular average over the $\Gamma_7$
``hot spots." The development
of the dip structure just below the Fermi level is correlated to
the resistivity bendover at low temperatures. This feature does not occur
in the two- and three-channel regimes.
A Kondo resonance related peak develops near
the Fermi energy, whose position with respect to the Fermi level
depends on the
channel character of the model parameters.
Details are displayed in Fig.~\ref{tau}.
Note that the anisotropic scattering
rate includes explicitly the hybridization as opposed to the other
spectral functions like the atomic electron spectral function and the
dynamic susceptibility.

 The transport coefficients are calculated using the Kubo formula under
the assumption of dominant scattering in the $l=3$ channel. Our results
are defined in terms of transport integrals $I_n$ given by
equations.
\bea
I_n (T) &=& \int d\omega ~ \omega^n ~ \tau (\omega, T) ~
     \left[ -{\partial f(\omega) \over \partial \omega} \right],
     \\
\tau (\omega, T)
 &\equiv& \int {d\hat{k} \over 4\pi} ~ \tau (\hat{k}, \omega, T).
\eea
Here $f(\omega)$ is Fermi function.
The resistivity is calculated using the equation:
\bea
 {1\over \rho (T) } &=& {ne^2 \over m} ~ I_0 (T).
\eea

As we stated before, the resistivity shows a bendover
at low temperatures in the one-channel regime within our simple model
(not shown). This derives from the depleted
spectral weight in the $f^1 - f^2$ sector and zeros of
$\Theta_7^{(5/2)} (\hat{k})$.

 In the two-channel and three-channel model parameter regimes,
the resistivity initially increases logarithmically through $T_0$
and then saturates with a power law to a constant
with further decreasing temperatures.
The resistivity near zero temperature obeys a scaling behavior as shown in
Fig.~\ref{res23ch} confirming our zero temperature analysis with the scaling
dimensions, $\Delta_2 = 1/2, \Delta_3 = 2/5$. These results
agree with the conformal field theory results\cite{cft,ludwig},
$\rho (T) = \rho(0) ~[~1- a [T/T_0]^{\Delta_n}~]$ for $T\le 0.06T_0$,
$~~\Delta_n = 2/(n+2)$
for the overcompensated multi-channel Kondo models (note that the power
law exponent is independent of the impurity spin size).
We note that the region where strict $T^{\Delta_n}$ behavior holds
is below about $0.05~T_0$. A fit to the resistivity of
Ce$_x$La$_{1-x}$Cu$_{2.2}$Si$_2$ (see Fig. \ref{resfit}) is good until
low temperatures where the data breaks below theory.
This suggests a possible crossover to a new fixed point which could be
set by intersite interactions (producing a spin molecular field)
or a weak noncubic symmetry for the Ce$^{3+}$ ions.

 The thermopower is a sensitive measure of the asymmetry in
the scattering rate and the density of states (DOS)
near the Fermi level. Since we are assuming the symmetrical
Lorentzian/Gaussian
DOS for the conduction band, the sign of the thermopower
is determined by the asymmetrical scattering rate.
\bea
 Q(T) &=& - {1\over eT} ~ {I_1 (T) \over I_0 (T) }.
\eea
As shown in Fig.~\ref{tep}, the thermopower shows explicitly
the distinct behavior for different numbers of channels
varying from the two-channel  regime,
through the three-channel , finally to the one-channel  regime.
The hybridization between $f^0$ and $f^1$ leads to the Kondo resonance
peak just above the Fermi level
and more spectral weight above the Fermi level.
Since particle scattering dominates, holes are the main carriers in the
one-channel cases, thus leading to a positive thermopower.
On the other hand, negative thermopower arises in
the $f^1 - f^2$ sector. In this case, hole scatterings
dominate and particles are the main carriers.
In the three-channel regime, energy structures are symmetric.
However, the double degeneracy in the $f^2$ configuration leads to
a weak hole scattering dominance resulting in a negative thermopower at low
temperatures. The overall magnitude of the thermopower is
slightly reduced when anisotropy at the cubic sites is included.

\section{Discussion and Conclusion}
 We have introduced and studied a realistic model Hamiltonian for Ce$^{3+}$
impurities with three configurations ($f^0$, $f^1$, $f^2$), which
are embedded in cubic normal metals. This simple model shows competition
between the Fermi liquid fixed point of the one-channel $S=1/2$ Kondo
model and the non-Fermi liquid fixed point of the two-channel $S=1/2$
Kondo model.

 We studied a simplified Anderson model using the NCA.
This simple model covers one-, two-, and three-channel Kondo
physics depending on the model parameters. All the calculated
physical quantities show the signatures of the Kondo effect
appropriate to the different numbers of relevant channels.
The magnetic susceptibility agrees with
the exact Bethe Ansatz results for the two- and three-channel model
parameter regime and has the correct scaling dimension agreeing
with the conformal field theory results.
Entropy and specific heat calculations also
agree with the Bethe ansatz results within ${\cal O} (1/N^2)$ approximation.
Though the resistivity in the one-channel regime bends over
with decreasing temperatures, we do not believe that this feature
will survive when all the energy level structures, especially
$f^1 J=5/2 \Gamma_8$, are included.
In the two- and three-channel parameter regimes,
the resistivity increases logarithmically
and saturates with decreasing temperatures. The low temperature
behavior leads to power laws in agreement with the conformal
field theory results.

Since the thermopower is very sensitive to the density of states
structure and the
scattering mechanism near the Fermi level, its sign and its magnitude
depend on which fixed point is stable at zero temperature.
In the one-channel regime, electrons are strongly scattered due to the
Kondo resonance above the Fermi level and the thermopower remains positive
definite and large due to the resonant scattering. In the two-channel case,
holes are scattered off the impurity sites stronger than electrons. Thus
electrons are the main carriers leading to the negative thermopower. In the
three-channel regime, though hole scattering is reduced compared to
the two-channel case, electrons are still the main carriers due to the
degeneracy imbalance between the singlet $f^0$ and the doublet
$f^2 J=4 \Gamma_3$.
The thermopower remains negative with reduced value.
We calculated the dynamic magnetic susceptibility and characterized it with
its peak position ($\Gamma (T)$) as a function of temperature.
We can see the clear difference between one-channel and the overcompensated
cases. While $\Gamma (T)$ decreases and saturates to a constant value of order
$T_K$ with decreasing temperatures in the one-channel case,
$\Gamma (T)$ goes to zero almost linearly with decreasing temperatures
in the two- and three-channel case.

 We now discuss the experimental relevance of our model study to the
Ce$_{1-x}$La$_x$Cu$_{2.2}$Si$_2$ alloy\cite{andraka}.
We already stressed the experimental evidences supporting
the two-channel Kondo effect in this alloy system in the section I.
The thermopower for CeCu$_{2}$Si$_2$ changes sign around 70 K and is
negative and large below\cite{chitep,ceybtep}.
As our numerical calculation shows, the thermopower is negative and
large in the two-channel regime. This result compares well with
the experimental findings for the stoichiometric system with $x=1$.
We believe the sign change comes from the Kondo resonance of
$f^0$ and $f^1 \Gamma_8$, which lies above the Fermi level.
Further experiments are required for the alloy system with excess Cu.
For comparison, we note that
CeAl$_2$ or CeAl$_3$\cite{ceybtep} have a positive thermopower
large compared to transition
metals at high temperatures and have a sign change at low temperature
which is still bigger than the Kondo temperature.
Our thermopower calculation and the thermopower dependence on the
unit cell volume\cite{jaccard} suggests that the alloy system
Ce$_{1-x}$La$_x$Cu$_{2.2}$Si$_2$ can go through the three-channel
model parameter regime with external pressure.
Renormalized atom calculations further suggest a destabilization of
$f^2$ relative to $f^0$ with initial increasing pressure
\cite{herbst}.

In addition to the thermopower,
the future neutron scattering experiments for
Ce$_{1-x}$La$_x$Cu$_{2.2}$Si$_2$ alloy
can search for the dependence of $\Gamma (T)$ (the peak position of the
dynamic magnetic susceptibility) consistent with our proposal for
the channel number.

\acknowledgments

This research was supported by a grant
from the U.S. Department of Energy, Office of Basic Energy Sciences,
Division of Materials
Research.  We thank Eunsik Kim for her careful reading of this paper
and thank L. N. Oliveira and J.W. Wilkins for
stimulating interactions.

\appendix

\section{NCA self energy in the presence of multiple, same irreps}

 In the $f^2$ configuration, there are 9 irreducible representations
of $\Gamma_3$ in a cubic symmetry. NCA self energy diagrams should be
generalized to get the right Kondo energy scale. As an example,
we may consider the mixing process between $f^1 J=5/2 \Gamma_7$ and
$f^2 \Gamma_3$'s. The relevant NCA integral equations are
\bea
\Sigma (\omega)
 &=& {2\over \pi} \sum_{ij} \sqrt{\Gamma_i \Gamma_j} ~ \int d\epsilon ~
    f(\epsilon) \tilde{N} (\epsilon) ~ D_{ij} (\epsilon + \omega),
   \\
\Pi_{ij} (\omega)
 &=& {2\over \pi} \sqrt{\Gamma_i \Gamma_j} ~ \int d\epsilon ~
    f(-\epsilon) \tilde{N} (\epsilon) ~ G (\epsilon + \omega).
\eea
The conduction electrons of only $\Gamma_8$ symmetry are
involved in the mixing process.
The $f^2 \Gamma_3$'s Green's function now becomes a $9 \times 9$
matrix. This generalization can be seen most clearly by looking at
the self energy diagrams of $f^2 \Gamma_3$'s. The incoming
$\Gamma_3$ does not need to be the same as the outgoing
$\Gamma_3$. This leads to the matrix Green's function for
$f^2 \Gamma_3$'s.

 According to the Schrieffer-Wolff transformation, the
effective Hamiltonian is
\bea
H_1 &=& J \sum_{n} \vec{S}_I \cdot \vec{S}_{cn} (0); ~~
 J = \sum_{i=1}^{9} {2 |V_{12}^{i}|^2 \over \epsilon_{2}^{i} - \epsilon_1 }.
\eea
Now we can show that the above NCA integral equations lead to
the right Kondo energy scale to leading order.
\bea
 G (\omega) &\to& {1\over \omega - \epsilon_1 + i\delta}.
\eea
With this replacement, the $\Gamma_3$ self energy matrix
becomes
\bea
\Pi_{ij} (\omega)
 &=& {2\over \pi} \sqrt{\Gamma_i \Gamma_j} ~ \log \left|
    {\omega - \epsilon_1 \over \omega - D - \epsilon_1 }
   \right|
\eea
for a symmetric flat conduction band DOS with a half width $D$.
Now the Kondo energy scale is determined by
\bea
\mbox{det} ~[~ (\omega - \epsilon_{2}^{i}) \delta_{ij}
     - \Pi_{ij} (\omega) ~] &=& 0.
\eea
With the substitution $\omega = \epsilon_1 - T_0$, we can find
the following
\bea
\log \left| {T_0 \over D + T_0} \right|
  &=& {1 \over N(0)J}, \\
N(0)J
 &=& {2\over \pi} ~ \sum_{i=1}^{9}
   {\Gamma_i \over \epsilon_{2}^{i} - \epsilon_1 }.
\eea
Here $\Gamma_i = \pi N(0) |V_{12}^{i}|^2$.

 In conclusion, we have shown that the inclusion of all the $\Gamma_3$'s
in the $f^2$ configuration leads to the enhanced two-channel
exchange coupling.

\section{Zero temperature analysis: overcompensated multichannel
Anderson model.}

We discuss the following NCA integral equations at zero temperature.
\bea
G_{\rm g} (z)
 &=& { 1 \over z - \epsilon_{\rm g} - \Sigma_{\rm g} (z) }; ~~
   \Sigma_{\rm g} (z)
  = { N_{\rm x} \Gamma \over \pi} \int d\epsilon \tilde{N} (-\epsilon)
     f(\epsilon) G_{\rm x} (z+\epsilon), \\
G_{\rm x} (z)
 &=& { 1 \over z - \epsilon_{\rm x} - \Sigma_{\rm x} (z) }; ~~
  \Sigma_{\rm x} (z) = {N_{\rm g} \Gamma \over \pi}
  \int d\epsilon \tilde{N} (\epsilon)
     f(\epsilon) G_{\rm g} (z+\epsilon).
\eea
Here $\tilde{N} (\epsilon)$ is the normalized DOS such that
$\tilde{N} (0)=1$. For definiteness, we assume that the excited
state has one less electron than the ground state.
Before diving into the zero temperature analysis, we point out
the applicability of NCA approach to the Anderson impurity model.
Whenever the ground and the excited states have degeneracies
of $N_{\rm g}$ and $N_{\rm x}$,  respectively,
the above form of NCA integral equations are obtained.
The above NCA integral equations also derive from the
$N_{\rm x}$-channel, $S_I=(N_{\rm g} -1)/2$ models.
This is an artifact of the NCA approach. The NCA can not distinguish
between these two different models.
Note that not all the Anderson models with the degeneracies of
$N_{\rm g}$ (the ground state) and $N_{\rm x}$ (the excited state)
map into the $N_{\rm x}$-channel, $S_I=(N_{\rm g} -1)/2$ models.
The Schrieffer-Wolff transformation is essential to see this
connection clearly. As an example, the excited triplets in the $f^2 J=4$
in our Ce$^{3+}$ model leads to the one-channel exchange interaction of the
impurity $f^1 \Gamma_7$ pseudo spin $S_I=1/2$ coupled to the $S_c=3/2$
conduction electrons\cite{kimcoxtobe}
instead of three-channel Kondo model.
With this restriction in mind, we now study the zero temperature
analysis of the above NCA integral equations\cite{coxruck,zero}.

The dynamical quantities we are interested in are
\bea
\rho (\omega)
 &=& \int d\epsilon ~
     [~ a_{\rm x} (\epsilon) A_{\rm g} (\epsilon + \omega) +
       A_{\rm x} (\epsilon) a_{\rm g} (\epsilon + \omega) ~], \\
\chi_{\rm g}^{''} (\omega)
 &=& \int d\epsilon ~[~a_{\rm g} (\epsilon) A_{\rm g} (\epsilon + \omega)
    - A_{\rm g} (\epsilon) a_{\rm g} (\epsilon + \omega)~], \\
a_{\rm g} (\omega)
 &\equiv& { e^{-\beta\omega} \over Z_{\rm f}} A_{\rm g} (\omega); ~~
  a_{\rm x} (\omega) \equiv { e^{-\beta\omega} \over Z_{\rm f}}
  A_{\rm x} (\omega).
\eea
We can show that
\bea
 a_{\rm g} (\omega) |G_{\rm g} (\omega)|^{-2}
  &=& { N_{\rm x} \Gamma \over \pi} \int d\epsilon \tilde{N} (-\epsilon)
     f(-\epsilon) a_{\rm x} (\omega+\epsilon), \\
 a_{\rm x} (\omega) |G_{\rm x} (\omega)|^{-2}
  &=& { N_{\rm g} \Gamma \over \pi} \int d\epsilon \tilde{N} (\epsilon)
     f(-\epsilon) a_{\rm g} (\omega+\epsilon).
\eea

At zero temperature, Fermi function is reduced to step function. Thus,
the self energy equations are simplified:
\bea
 \Sigma_{\rm g} (z)
   &=& { N_{\rm x} \Gamma \over \pi}
    \int_{-\infty}^{0} d\epsilon \tilde{N} (-\epsilon)
    G_{\rm x} (z+\epsilon), \\
 \Sigma_{\rm x} (z)
   &=& {N_{\rm g} \Gamma \over \pi}
    \int_{-\infty}^{0} d\epsilon \tilde{N} (\epsilon)
    G_{\rm g} (z + \epsilon), \\
 a_{\rm g} (\omega) |G_{\rm g} (\omega)|^{-2}
  &=& { N_{\rm x} \Gamma \over \pi} \int_0^{\infty} d\epsilon ~
     \tilde{N} (-\epsilon) a_{\rm x} (\omega+\epsilon), \\
 a_{\rm x} (\omega) |G_{\rm x} (\omega)|^{-2}
  &=& { N_{\rm g} \Gamma \over \pi} \int_0^{\infty} d\epsilon ~
     \tilde{N} (\epsilon) a_{\rm g} (\omega+\epsilon).
\eea
For a flat conduction band with cut-off, $[-D, D]$, we can reduce the above
equations to differential equations:
\bea
 g_{\rm g} (\omega)
  &=& - 1/G_{\rm g} (\omega); ~~ g_{\rm x} (\omega)
     = - 1/G_{\rm x} (\omega), \\
 {d \over d\omega} g_{\rm g} (\omega)
  &=& -1 - { N_{\rm x} \Gamma \over \pi} { 1 \over g_{\rm x} (\omega) };
    ~~ g_{\rm g} (-D) = D + \epsilon_{\rm g}, \\
 {d \over d\omega} g_{\rm x} (\omega)
  &=& -1 - { N_{\rm g} \Gamma \over \pi} { 1 \over g_{\rm g} (\omega) };
    ~~ g_{\rm x} (-D) = D + \epsilon_{\rm x}, \\
 {d \over d\omega} \left[ a_{\rm g} (\omega) |g_{\rm g} (\omega)|^{2}
    \right]
  &=& - { N_{\rm x} \Gamma \over \pi} a_{\rm x} (\omega), \\
 {d \over d\omega} \left[ a_{\rm x} (\omega) |g_{\rm x} (\omega)|^{2}
    \right]
  &=& - { N_{\rm g} \Gamma \over \pi} a_{\rm g} (\omega).
\eea
Removing the $\omega$ dependence, we find the relationship between $g_{\rm g}$
and $g_{\rm x}$, in terms of the integration constant which connects the low
energy and
high energy states.
\bea
{g_{\rm g} \over D + \epsilon_{\rm g} }
 &=& \exp \left[{\pi (\epsilon_{\rm g} - \epsilon_{\rm x})
      \over N_{\rm g} \Gamma} \right]
   \exp \left[-{\pi (g_{\rm g} - g_{\rm x}) \over N_{\rm g} \Gamma} \right]
   \left[ {g_{\rm x} \over D + \epsilon_{\rm x} }
   \right]^{N_{\rm x}/N_{\rm g}}.
\eea
Since the zero temperature analysis is meaningful only when $D \gg
\epsilon_{\rm g, x}$, we will replace $D + \epsilon_{\rm g, x}$
by $D$. That is,
\bea
{g_{\rm g} \over T_0 }
 &=& \exp \left[-{\pi (g_{\rm g} - g_{\rm x}) \over N_{\rm g} \Gamma} \right]
   \left[ {g_{\rm x} \over \Delta } \right]^{N_{\rm x}/N_{\rm g}},
   \\
T_0 &=& D ~\left[\Delta \over D \right]^{N_{\rm x} / N_{\rm g}}~
   \exp \left[{\pi (\epsilon_{\rm g} - \epsilon_{\rm x})
      \over N_{\rm g} \Gamma} \right], \\
\Delta &=& {\Gamma \over \pi}.
\eea
And
\bea
 {d \over d\omega} ~[~ N_{\rm g} a_{\rm g} (\omega) g_{\rm g} (\omega)
  + N_{\rm x} a_{\rm x} (\omega) g_{\rm x} (\omega) ~]
  &=& N_{\rm g} a_{\rm g} (\omega) + N_{\rm x} a_{\rm x} (\omega).
\eea
Due to the sharpness of Fermi function at the Fermi level, the spectral
functions have a sharp cut-off at the threshold energy $E_0$.
Since $A_n(\omega)$ vanishes below this cut-off energy, $\Sigma_n (\omega)$
is purely real and $g_n (\omega)$ does not vanish or cross the frequency
axis below the threshold energy $E_0$. That is, $g_n (\omega)$ is positive
definite below the threshold energy $E_0$.

\subsection{Leading asymptotic behavior}
 We can derive the asymptotic behavior near the
cut-off energy. Since $a_{\rm g} (\omega)$ and $g_{\rm x} (\omega)$ vanish at
$\omega = E_0$, we can approximate the above equations:
\bea
E_0 - \omega
  &=& \int_0^{g_{\rm x}} dy { g_{\rm g} (y) \over g_{\rm g} (y)
   + N_{\rm g} \Delta }
  \approx {1 \over N_{\rm g} \Delta } \int_0^{g_{\rm x}} dy ~ g_{\rm g} (y),
    \\
E_0 - \omega
  &=& \int_0^{g_{\rm g}} dy { g_{\rm x} (y) \over g_{\rm x} (y)
   + N_{\rm x} \Delta }
  \approx {1 \over N_{\rm x} \Delta } \int_0^{g_{\rm g}} dy ~ g_{\rm x} (y),
   \\
{g_{\rm g} \over T_0}
  &\approx& \left[{g_{\rm x} \over \Delta} \right]^{N_{\rm x}/N_{\rm g}}.
\eea
Here $\Delta = \Gamma / \pi$. From the above, we can find
\bea
 { g_{\rm g} (\omega) \over T_0 }
  &\approx& ~ |\tilde{\Omega}|^{\alpha_{\rm g}}, \\
 { g_{\rm x} (\omega) \over \Delta }
  &\approx& ~ |\tilde{\Omega}|^{\alpha_{\rm x}}, \\
 T_0 &=& D \left[{\Delta \over D}\right]^{N_{\rm x}/N_{\rm g}}
     \exp \left[{\pi (\epsilon_{\rm g} - \epsilon_{\rm x})
      \over N_{\rm g} \Gamma} \right], \\
 \alpha_{\rm g} &=& { N_{\rm x} \over N_{\rm g} + N_{\rm x}}; ~~
  \alpha_{\rm x} = { N_{\rm g} \over N_{\rm g} + N_{\rm x}}, \\
\tilde{\Omega}
 &\equiv& (N_{\rm g} + N_{\rm x}) { E_0 - \omega \over T_0 }.
\eea
Since the zero temperature analysis is based upon the assumption
$|\epsilon_n| \ll D$, the realistic Kondo energy scale is given by
the replacement of $D+\epsilon_n \rightarrow D$.
The asymptotic behavior right above the cut-off $E_0$ can be found from the
the expressions below $E_0$ by the analytic continuation:
\bea
 { g_{\rm g} (\omega + i\delta) \over T_0 }
  &\approx& e^{-i\alpha_{\rm g} \pi}
  ~ \tilde{\Omega}^{\alpha_{\rm g}},\\
 { g_{\rm x} (\omega + i\delta) \over \Delta }
  &\approx& e^{-i\alpha_{\rm x} \pi}
  ~ \tilde{\Omega}^{\alpha_{\rm x}}, \\
 A_{\rm g} (\omega)
  &\approx& {1 \over \pi T_0} \sin(\alpha_{\rm g} \pi)
   ~ |\tilde{\Omega}|^{-\alpha_{\rm g}} \theta (\omega - E_0), \\
 A_{\rm x} (\omega)
  &\approx& {1 \over \pi \Delta} \sin(\alpha_{\rm x} \pi)
   ~ |\tilde{\Omega}|^{-\alpha_{\rm x}} \theta (\omega - E_0).
\eea
Here the phase was determined such that the spectral function is positive
definite above the threshold energy.
Since $[~ N_{\rm g} a_{\rm g} g_{\rm g} + N_{\rm x} a_{\rm x} g_{\rm x}
{}~]_{\omega = E_0} =1$, we can deduce that
\bea
 a_{\rm g} (\omega)
  &\approx& { 1 \over (N_{\rm g} + N_{\rm x}) T_0 }
  ~ \tilde{\Omega}^{-\alpha_{\rm g}} \theta (E_0 - \omega), \\
 a_{\rm x} (\omega)
  &\approx& { 1 \over (N_{\rm g} + N_{\rm x}) \Delta }
  ~ \tilde{\Omega}^{-\alpha_{\rm x}} \theta (E_0 - \omega).
\eea
{}From the above asymptotic expressions, we find
\bea
\rho (\omega)
 &\approx& { 1 \over \Gamma } { 1 \over (N_{\rm g} + N_{\rm x})^2}
     ~ B(\alpha_{\rm x},\alpha_{\rm g}) ~
  \cases{ \sin(\alpha_{\rm g} \pi) ~ \theta(\omega) \cr
          \sin(\alpha_{\rm x} \pi)~ \theta(-\omega) \cr} \nonumber\\
 &=& {1\over \Delta} ~ {1 \over [N_{\rm g} + N_{\rm x}]^2}, \\
\chi_{\rm g}^{''} (\omega)
 &\approx& { 1 \over \pi T_0 }
   { \sin(\alpha_{\rm g} \pi) \over (N_{\rm g} + N_{\rm x})^2}
   ~ B (\alpha_{\rm x}, \alpha_{\rm x}) ~ \mbox{sgn}(\omega) ~
   |\tilde{\omega} |^{1-2\alpha_{\rm g}}, \\
\tilde{\omega}
 &\equiv & (N_{\rm g} + N_{\rm x}) { \omega \over T_0 }.
\eea
Here $B(p,q)$ is the beta function.

\subsection{Next leading asymptotic behavior}
{}From the above analysis, we can see that it is more appropriate to
use the dimensionless quantities. Here we collect all the relevant formulas
from the above.
\bea
{d \over d\tilde{\Omega}} {g_{\rm g} \over T_0}
 &=& \alpha_{\rm g} \left[ {1\over N_{\rm x} }
                           + {\Delta \over g_{\rm x} } \right];
    ~~ g_{\rm g} (-D) = D + \epsilon_{\rm g}, \\
{d \over d\tilde{\Omega}} {g_{\rm x} \over \Delta}
 &=& \alpha_{\rm x} \left[ {T_0 \over N_{\rm g} \Delta}
         + {T_0 \over g_{\rm g}} \right];
    ~~ g_{\rm x} (-D) = D + \epsilon_{\rm x}, \\
{d \over d\tilde{\Omega}} \left[ a_{\rm g} (\omega) |g_{\rm g} (\omega)|^{2}
    \right]
  &=& \alpha_{\rm g} \Delta~T_0 ~ a_{\rm x} (\omega), \\
{d \over d\tilde{\Omega}} \left[ a_{\rm x} (\omega) |g_{\rm x} (\omega)|^{2}
    \right]
  &=& \alpha_{\rm x} \Delta~T_0 ~ a_{\rm g} (\omega), \\
E_0 - \omega
 &=& \int_0^{g_{\rm x}} dy { g_{\rm g} (y) \over g_{\rm g} (y)
   + N_{\rm g} \Delta }, \\
E_0 - \omega
 &=& \int_0^{g_{\rm g}} dy { g_{\rm x} (y) \over g_{\rm x} (y)
   + N_{\rm x} \Delta }, \\
{g_{\rm g} \over T_0}
 &=& \exp \left[-{\pi (g_{\rm g} - g_{\rm x}) \over N_{\rm g} \Gamma} \right] ~
   ~\left[ {g_{\rm x} \over \Delta} \right]^{N_{\rm x}/N_{\rm g}}.
\eea
Expanding the last relation, we get
\bea
{g_{\rm g} \over T_0}
 &=& \left[ {g_{\rm x} \over \Delta} \right]^{N_{\rm x}/N_{\rm g}} ~
  \left( 1 + {g_{\rm x} \over N_{\rm g} \Delta}
     - {T_0 \over N_{\rm g} \Delta}~
   \left[{g_{\rm x} \over \Delta}\right]^{N_{\rm x} / N_{\rm g}} + \cdots
  \right).
\eea
Then it is straightforward to show that
\bea
{g_{\rm g} \over T_0}
 &=& \tilde{\Omega}^{\alpha_{\rm g}} ~ \left[
    1 - g_1 ~ \tilde{\Omega}^{\alpha_{\rm g}}
    + 2g_2 ~ \tilde{\Omega}^{\alpha_{\rm x}} + \cdots ~\right], \\
{g_{\rm x} \over \Delta}
 &=& \tilde{\Omega}^{\alpha_{\rm x}} ~ \left[
    1 + 2g_1 ~ \tilde{\Omega}^{\alpha_{\rm g}}
    - g_2 ~ \tilde{\Omega}^{\alpha_{\rm x}} + \cdots ~\right], \\
g_1 &=& {1\over N_{\rm g} + 2N_{\rm x} } ~ { T_0 \over \Delta }; ~~
 g_2 = {1\over 2N_{\rm g} + N_{\rm x} }.
\eea
Furthermore, writing in the Taylor expansion form,
\bea
a_{\rm g}
 &=& {1 \over [~N_{\rm g} + N_{\rm x}~]~T_0}~
    \tilde{\Omega}^{-\alpha_{\rm g}} ~
  \left[1 + a_1 ~ \tilde{\Omega}^{\alpha_{\rm g}}
    + a_2 ~ \tilde{\Omega}^{\alpha_{\rm x}} + \cdots \right]
  \theta (\tilde{\Omega}), \\
a_{\rm x}
 &=& {1 \over [~N_{\rm g} + N_{\rm x}~]~\Delta}~
    \tilde{\Omega}^{-\alpha_{\rm x}} ~
  \left[1 + b_1 ~ \tilde{\Omega}^{\alpha_{\rm g}}
    + b_2 ~ \tilde{\Omega}^{\alpha_{\rm x}} + \cdots \right]
  \theta (\tilde{\Omega}).
\eea
we can find the following relations
\bea
&& b_1 = 2a_1 - 4g_1; ~~ \alpha_{\rm g} b_2 = a_2 + 4g_2, \\
&& \alpha_{\rm x} a_1 = b_1 + 4g_1; ~~ a_2 = 2b_2 - 4g_2.
\eea
Finally we find
\bea
a_{\rm g}
 &=& {1 \over [~N_{\rm g} + N_{\rm x}~]~T_0}~
   \tilde{\Omega}^{-\alpha_{\rm g}} ~
  \left[1 + 0 ~\tilde{\Omega}^{\alpha_{\rm g}}
        - 4g_2 ~ \tilde{\Omega}^{\alpha_{\rm x}} + \cdots \right]
  \theta (\tilde{\Omega}), \\
a_{\rm x}
 &=& {1 \over [~N_{\rm g} + N_{\rm x}~]~\Delta}~
   \tilde{\Omega}^{-\alpha_{\rm x}} ~
  \left[1 - 4g_1 ~ \tilde{\Omega}^{\alpha_{\rm g}}
    + 0 ~ \tilde{\Omega}^{\alpha_{\rm x}} + \cdots \right]
  \theta (\tilde{\Omega}).
\eea
And the pseudo spectral functions are
\bea
A_{\rm g}
 &=& {1 \over \pi T_0}~ |\tilde{\Omega}|^{-\alpha_{\rm g}} ~
  \left[ \sin (\alpha_{\rm g} \pi)
    - 2g_2 ~\sin [(\alpha_{\rm g}-\alpha_{\rm x}) \pi] ~
       |\tilde{\Omega}|^{\alpha_{\rm x}} + \cdots \right]
    \theta (-\tilde{\Omega}) \nonumber\\
 &=& {\sin (\alpha_{\rm g} \pi) \over \pi T_0}~
     |\tilde{\Omega}|^{-\alpha_{\rm g}} ~
  \left[ 1 + 4g_2 ~\cos (\alpha_{\rm g} \pi) ~
       |\tilde{\Omega}|^{\alpha_{\rm x}} + \cdots \right]
    \theta (-\tilde{\Omega}), \\
A_{\rm x}
 &=& {\sin (\alpha_{\rm g} \pi) \over \pi \Delta}~
     |\tilde{\Omega}|^{-\alpha_{\rm x}} ~
  \left[ 1 - 4g_1 ~\cos (\alpha_{\rm g} \pi) ~
       |\tilde{\Omega}|^{\alpha_{\rm g}} + \cdots \right]
    \theta (-\tilde{\Omega}).
\eea
The scaling dimensions found here all agree with those found in the conformal
field theory for the overcompensated cases.

\subsection{Physical quantities}
Using the above results, we can find the dynamic susceptibilities for the
ground and the excited configurations
\bea
\chi_{\rm g}^{''} (\omega)
 &=& \mbox{sgn} (\omega) ~
  {\sin(\alpha_{\rm g} \pi) \over (N_{\rm g} + N_{\rm x})^2 \pi T_0}
   \nonumber\\
 && \times
  \left[ B(\alpha_{\rm x}, \alpha_{\rm x}) ~ |\tilde{\omega}|^{1-2\alpha_{\rm
g}}
   - 4g_2~[~1 - \cos(\alpha_{\rm g} \pi)~]~ B(\alpha_{\rm x}, 2\alpha_{\rm x})
    ~ |\tilde{\omega}|^{2-3\alpha_{\rm g}} + \cdots \right], \\
\chi_{\rm x}^{''} (\omega)
 &=& \mbox{sgn} (\omega) ~
  {T_0 \sin(\alpha_{\rm x} \pi) \over (N_{\rm g} + N_{\rm x})^2 \pi \Delta^2} ~
    \nonumber\\
 && \times
  \left[ B(\alpha_{\rm g}, \alpha_{\rm g}) ~ |\tilde{\omega}|^{1-2\alpha_{\rm
x}}
   - 4g_1 ~[~1 - \cos(\alpha_{\rm x} \pi)~]~ B(\alpha_{\rm g}, 2\alpha_{\rm g})
    ~ |\tilde{\omega}|^{2-3\alpha_{\rm x}} + \cdots \right].
\eea
These correlation functions are directly proportional to the imaginary part of
the corresponding dynamic correlation functions. The functional forms are
marginal Fermi liquid type for the case $N_{\rm g} = N_{\rm x}$,
which correspond to the overcompensated case.
The convoluted atomic electron spectral functions are
\bea
\rho_{\rm f} (\omega > 0)
 &=& {\sin(\alpha_{\rm g} \pi) \over (N_{\rm g} + N_{\rm x})^2 \pi \Delta}
   \nonumber\\
 && \times
  \left[ B(\alpha_{\rm x}, \alpha_{\rm g})
   - 4 g_1 ~ B(\alpha_{\rm x}, 2\alpha_{\rm g})
    ~ |\tilde{\omega}|^{\alpha_{\rm g}}
   + 4g_2~ \cos(\alpha_{\rm g} \pi)~ B(\alpha_{\rm g}, 2\alpha_{\rm x})
          |\tilde{\omega}|^{\alpha_{\rm x}} + \cdots \right], \\
\rho_{\rm f} (\omega < 0)
 &=& {\sin(\alpha_{\rm x} \pi) \over (N_{\rm g} + N_{\rm x})^2 \pi \Delta}
   \nonumber\\
 && \times
  \left[ B(\alpha_{\rm x}, \alpha_{\rm g})
   + 4g_1~ \cos(\alpha_{\rm x} \pi)~ B(\alpha_{\rm x}, 2\alpha_{\rm g})
          |\tilde{\omega}|^{\alpha_{\rm g}}
   - 4 g_2 ~ B(\alpha_{\rm g}, 2\alpha_{\rm x})
    ~ |\tilde{\omega}|^{\alpha_{\rm x}} + \cdots \right].
\eea
One important observation is that the scaling dimensions agree with those
obtained from the conformal field theory treatments for the overcompensated
multichannel $S=1/2$ models. From this result, we can argue
that the convoluted local electron spectral function is peaked right
at the Fermi level (Kondo resonance) for the overcompensated models.
This result seems to be independent of the occupancy of the
ground level. In fact, the numerical results with NCA confirm this
conclusion. Hence we conclude that Kondo resonance peak
in the overcompensated models sits right at the Fermi level irrespective
of any model parameters. In the single channel Anderson model, the position
of the Kondo resonance peak is adjusted by the Friedel sum rule (Fermi liquid
ground state). That is, the occupancy of the ground configuration determines
the Kondo resonance peak position.

We can also generate the low temperature dependence of some physical
quantities. Resistivity and thermopower can be evaluated using
the Kubo formula.
\bea
\rho (T)
 &=& \rho (0) \left( 1 - c \left[{T\over T_0}\right]^{\nu} \right); ~~
   \nu = \mbox{min} (\alpha_{\rm g}, \alpha_{\rm x}), \\
Q(T) &\propto& \left[{T\over T_0}\right]^{\nu}.
\eea

\section{Conduction electron scattering rate.}

 Here we derive the conduction electron scattering time in the presence
of the Anderson magnetic impurities.
In general, the conduction electron scattering rate (the
inverse of the conduction electron lifetime) is different from the
transport scattering rate. This difference derives from the vertex correction
when we calculate the current-current response function. The simplifying
feature of the Anderson model is that only one partial wave state
is coupled to the atomic electron state for an isotropic hybridization.
In this simplified model, the transport scattering rate is proportional
to the imaginary part of the conduction electron self energy.

The conduction electron scattering time is given
by the angular average of the anisotropic one.
\bea
\tau_{\mu\nu} (\omega)
 &=& 3 \int {d\hat{k} \over 4\pi} ~ \hat{k}_{\mu} \hat{k}_{\nu} ~
     \tau_{\mu\nu} (\hat{k}, \omega).
\eea
Since $\tau_{\mu\nu} (-\hat{k}, \omega) = \tau_{\mu\nu} (\hat{k}, \omega)$
in the Anderson model, the scattering time matrix becomes diagonal.
Thus we have
\bea
\tau_{\mu\mu} (\omega)
 &=& \int {d\hat{k} \over 4 \pi} ~\hat{k}_{\mu} \hat{k}_{\mu} ~
   \tau (\hat{k}, \omega).
\eea
In the dilute Anderson impurity limit, the conduction electron self energy
can be approximated by\cite{bcw}
\bea
\Sigma (\vec{k}\alpha, i\omega)
 &=& N_{\rm imp} \sum_i |< \vec{k}\alpha |V| i>|^2 ~ G_i (i\omega),
   \\
\tau^{-1} (\vec{k}\alpha, \omega)
 &=& 2 \pi N_{\rm imp} \sum_i |< \vec{k}\alpha |V| i>|^2 ~ \rho_i (\omega).
\eea
Here the index $i$ is the atomic electron's good quantum number.
We can write the self energy as
\ben
\tm With LS coupling, but without CEF.
\bea
\Sigma (\vec{k}\alpha, i\omega)
 &=& N_{\rm imp} \sum_{jm} |< \vec{k}\alpha |V| jm >|^2
   ~ G_{\rm f} (j; i\omega), \\
G_{\rm f} (j; \tau)
 &=& - < T_{\tau} f_{jm} (\tau) f_{jm}^{\dagger} (0) >, \\
< \vec{k}\alpha |V| jm >
 &=& \sum_{m_3 \beta} < \vec{k}\alpha |V| m_3 \beta >
     ~ < 3 m_3; {1\over 2} \beta | jm >,
\eea
The total angular momentum is a good quantum number in this case.

\tm With both LS coupling and CEF.
\bea
\Sigma (\vec{k}\alpha, i\omega)
 &=& N_{\rm imp} \sum_{j c d_c}
    |< \vec{k}\alpha |V| \Gamma_c^{(j)} d_c>|^2
   ~ G_{\rm f} ( \Gamma_c^{(j)}; i\omega),
   \\
G_{\rm f} (\Gamma_c^{(j)}; \tau)
 &=& - < T_{\tau} f_{\Gamma_c^{(j)} d_c} (\tau)
       f_{\Gamma_{c}^{(j)} d_{c}}^{\dagger} (0) >, \\
< \vec{k}\alpha |V| \Gamma_c^{(j)} d_c >
 &=& \sum_{m} < \vec{k}\alpha |V| jm >
    ~ <jm|\Gamma_c^{(j)} d_c>.
\eea
The CEF irreps are good quantum numbers.

\een
In the above, two successive unitary transformations have been used.
\bea
f_{m_3 \alpha}
 &=& \sum_{jm} <3 m_3; {1\over 2} \alpha | jm> ~ f_{jm} \\
f_{jm} &=& \sum_{c d_c} <jm|\Gamma_c^{(j)} d_c> ~ f_{\Gamma_c^{(j)} d_c}.
\eea

For the isotropic, spin-independent hybridization and the free electron
conduction band, the mixing matrix is given by
\bea
< \vec{k}\alpha |V| m_3 \beta >
 &=& \sqrt{4\pi \over \Omega}~V (k)Y_{3 m_3}(\hat{k})~\delta_{\alpha \beta}
    \\
V (k) &=& \sqrt{4\pi} ~ (-i)^3 \int dr ~ r^2 j_3 (kr) V(r) R_{\rm f} (r).
   \\
< \vec{k}\alpha |V| jm >
 &=& \sqrt{4\pi \over \Omega} ~ V (k) <\hat{k} \alpha | jm> \\
<\hat{k} \alpha | jm>
 &=& \sum_{m_3} Y_{3 m_3} (\hat{k}) ~ < 3 m_3; {1\over 2} \alpha | jm >
   \\
< \vec{k}\alpha |V| \Gamma_c^{(j)} d_c >
 &=& \sqrt{4\pi \over \Omega} ~ V (k) <\hat{k} \alpha | \Gamma_c^{(j)} d_c >
  \\
<\hat{k} \alpha | \Gamma_c^{(j)} d_c >
 &=& \sum_{m_3 m} Y_{3 m_3} (\hat{k}) ~ < 3 m_3; {1\over 2} \alpha | jm >
   <jm| \Gamma_c^{(j)} d_c >.
\eea
Here $\Omega$ is the volume of the system.
Hence the conduction electron self energy in the isotropic hybridization
can be written as
\bea
\Sigma (\vec{k}\alpha, i\omega)
  &=& 4 \pi n_{\rm imp} ~|V(k)|^2 \sum_{jm} |<\hat{k} \alpha | jm>|^2 ~
  ~ G_{\rm f} (j; i\omega), \\
  &=& 4 \pi n_{\rm imp} ~|V(k)|^2 \sum_{jc d_c}
   |<\hat{k} \alpha | \Gamma_c^{(j)} d_c>|^2 ~
  ~ G_{\rm f} (\Gamma_c^{(j)}; i\omega).
\eea
Here only the diagonal elements of Green's function are nonvanishing.

\subsection{Without CEF}
 When the CEF is neglected, the conduction electron self energy in a dilute
impurity limit reads
\bea
\Sigma (\vec{k}\alpha, i\omega)
  &=& 4 \pi n_{\rm imp} | V(k) |^2 \sum_{j}
     \Theta^{(j)} (\hat{k}) ~ G_{\rm f} (j; i\omega), \\
\tau^{-1} (\vec{k}\alpha, \omega)
  &=& 8 \pi^2 n_{\rm imp} | V(k) |^2 \sum_{j}
     \Theta^{(j)} (\hat{k}) ~ \rho_{\rm f} (j; \omega), \\
\Theta^{(j)} (\hat{k})
 &\equiv& \sum_{m} |< \hat{k}\alpha |jm >|^2.
\eea
$\rho_{\rm f} (j; \omega)$ is the measurable spectral function for the
atomic electrons with the total angular momentum $j$.

 The relevant angular functions defined above are, in fact, constants.
\bea
\Theta^{(j)} (\hat{k})
 &=& \sum_{m} |< \hat{k}\alpha |jm >|^2
  = \sum_{\mu d_{\mu}} |< \hat{k}\alpha |\Gamma_{\mu}^{(j)} d_{\mu} >|^2
  = {2j+1 \over 8\pi}.
\eea

\subsection{With CEF}
 In the presence of the CEF, the CEF irreducible representations are good
quantum numbers. The atomic $f$ electron operator can be decomposed into
CEF irreducible representation components.
Hence the conduction electron self energy in the dilute impurity limit is
\bea
\Sigma (\vec{k}\alpha, i\omega)
  &=& 4 \pi n_{\rm imp} ~|V(k)|^2 \sum_{jc}
   \Theta_{c}^{(j)} (\hat{k}) ~ G_{\rm f} (\Gamma_c^{(j)}; i\omega), \\
\Theta_{\Gamma_c}^{(j)} (\hat{k})
 &\equiv& \sum_{d_c} | <\hat{k} \alpha | \Gamma_c^{(j)} d_c> |^2.
\eea
Here $\Theta_{\Gamma_c}^{(j)} (\hat{k})$ is crystal harmonics.
The anisotropic relaxation rate reads
\bea
\tau^{-1} (\vec{k}\alpha, \omega)
  &=& 8 \pi^2 n_{\rm imp} ~|V(k)|^2 \sum_{jc}
   \Theta_{\Gamma_c}^{(j)}(\hat{k})~\rho_{\rm f}(\Gamma_c^{(j)}; \omega).
\eea
Here $\rho_{\rm f} (\Gamma_c^{(j)}; \omega)$ is the measurable
spectral function for the atomic $f$ electron of $\Gamma_{c}^{(j)}$
symmetry. This spectral function is given by the convolution of two
neighboring configuration Green's functions.

 In the cubic crystal symmetry, the relevant crystal angular functions are
\bea
\Theta_7^{(5/2)} (\hat{k})
 &=& \sum_{\alpha} |<\hat{k} \up / \down| \Gamma_{7}^{(5/2)} \alpha > |^2
     \nonumber\\
 &=& - { 1\over 32 \pi} ~[~ 35\cos^4 \theta  - 30 \cos^2 \theta - 5
  + 5 \sin^4 \theta \cos 4\varphi ~], \\
\Theta_8^{(5/2)} (\hat{k})
 &=& \sum_{n\alpha} |<\hat{k} \up / \down| \Gamma_{8}^{(5/2)}; n \alpha > |^2
   \nonumber\\
 &=& { 1\over 32 \pi} ~[~ 35\cos^4 \theta  - 30 \cos^2 \theta + 19
  + 5 \sin^4 \theta \cos 4\varphi ~].
\eea
And
\bea
\Theta_6^{(7/2)} (\hat{k})
 &=& \sum_{\alpha} |<\hat{k} \up / \down| \Gamma_{6}^{(7/2)}; \alpha > |^2
    \nonumber\\
 &=& { 7\over 256 \pi} ~[~ 50 \cos^6 \theta - 30 \cos^4 \theta
  - 10 \cos^2 \theta + {34\over 3} \nonumber\\
 && \hspace{1.0cm} - 10 \sin^4 \theta (5 \cos^2 \theta - 1) \cos 4\varphi ~],
\\
\Theta_7^{(7/2)} (\hat{k})
 &=& \sum_{\alpha} |<\hat{k} \up / \down| \Gamma_{7}^{(7/2)}; \alpha > |^2
    \nonumber\\
 &=& { 15\over 256 \pi} \sin^2 \theta
   ~[~ - 14 \cos^4 \theta  + 28 \cos^2 \theta + 2 \nonumber\\
 && \hspace{1.0cm} - 2 \sin^2 \theta (7 \cos^2 \theta + 1) \cos 4\varphi ~], \\
\Theta_8^{(7/2)} (\hat{k})
 &=& \sum_{n\alpha} |<\hat{k} \up / \down| \Gamma_{8}^{(7/2)}; n \alpha >|^2
    \nonumber\\
 &=& { 1\over 32 \pi} ~[~ -70 \cos^6 \theta + 105 \cos^4 \theta
  - 40 \cos^2 \theta + {55 \over 3} \nonumber\\
 && \hspace{1.0cm} + 5 \sin^4 \theta (14 \cos^2 \theta -1) \cos 4\varphi ~].
\eea
Note that the above crystal angular functions are normalized and satisfy
the correct sum rules.
\bea
\int d\hat{k} ~ \Theta_{6,7}^{(j)} (\hat{k})
 &=& 1; ~~ \int d\hat{k} ~ \Theta_{8}^{(j)} (\hat{k}) =2, \\
\sum_{\mu}  \Theta_{\mu}^{(j)} (\hat{k})
 &=& {2j+1 \over 8\pi}.
\eea
Furthermore, the crystal harmonics for $j=5/2$ multiplets can be rewritten as
\bea
\Theta_7^{(5/2)} (\hat{k})
 &=& {1\over 16\pi} ~[~ 6 - \Phi (\hat{k}) ~]; ~~
    \Theta_8^{(5/2)} (\hat{k}) = {1\over 16\pi}
     ~[~ 6 + \Phi (\hat{k}) ~], \\
\Phi (\hat{k})
 &=& 15 \cos^4 \theta - 10 \cos^2 \theta + 1
    + 5\sin^4 \theta \cos^2 2\varphi.
\eea

\begin{figure}
\vskip 1.0cm
\protect\caption[Fig.1.]
{{\bf Comparison of experimental resistivity with our numerical
calculation.}
Our numerical calculation shows a $T^{1/2}$ behavior
at low temperatures. Experimental resistivity (points
with $T_0=10$ K, from Ref.\cite{andraka})
is compared with our numerical results.
Three different symbols mean a set of three model parameters
($\Box$ for model set 1; $\bigcirc$ for model set 2;
$\triangle$ for model set 3).
The low temperature deviation suggests a possible crossover to
a new fixed point.}
\protect\label{resfit}
\end{figure}

\newpage
\begin{figure}
\vskip 1.0cm
\protect\caption[Fig.1.]
{{\bf Crystal electric field energy level scheme
for $f^0, f^1, f^2$ configurations.}
The one-channel \& two-channel Anderson model Hamiltonians are developed from
these CEF energy states. The one-channel Kondo model derives from the $f^0$
singlet and the magnetic $f^1 \Gamma_7$ doublet which mix through hybridization
with the $\Gamma_7$ conduction electrons.
The two-channel Kondo model derives from the magnetic $f^1 \Gamma_7$ doublet
and the nonmagnetic $f^2 \Gamma_3$ doublet which mix through hybridization
with the $\Gamma_8$ conduction electrons.}
\protect\label{cef}
\end{figure}

\newpage
\begin{figure}
\vskip 0.5cm
\protect\caption[Fig.1.]
{{\bf Schematic digrams of two channel degrees of freedom.} The upper
figure is for $\Gamma_8 +$ and the lower figure for $\Gamma_8 -$
of $J=5/2$ conduction electron partial waves.  }
\protect\label{channel}
\end{figure}

\newpage
\begin{figure}
\protect\caption[Fig.1.]
{{\bf Scaling diagrams of one-channel \& two-channel Anderson model
up to third order.} Dashed lines are for the Ce$^{3+}$
impurity $\Gamma_7$ pseudo spins and the solid lines are for the $\Gamma_7$
or $\Gamma_{8n}$ conduction electrons. The diagrams labeled as above lead
to the scaling equation for the one-channel exchange coupling, $J_1$.
To obtain the scaling equations for the two-channel exchange coupling,
$J_2$, the labels $\Gamma_7$ for the external conduction electrons (solid
lines)
should be replaced by $\Gamma_{8n}$. }
\protect\label{scale}
\end{figure}

\newpage
\begin{figure}
\vskip 1.0cm
\protect\caption[Fig.1.]
{{\bf Leading skeleton self energy diagrams from the NCA.}
Since two different
symmetry conduction electrons are involved in the hybridizations
of $f^0-f^1$ and $f^1-f^2$, our NCA self energy diagrams become
simplified. The diagram (a) is the self energy for the
$f^0$ atomic state (wiggly line). The diagrams (b) are
for the $f^1 \Gamma_7$ atomic state (dashed line). The diagram (c) is
for the $f^2 \Gamma_3$ atomic state (dotted line).
The solid line is the conduction electron
propagator: the first two are for the $\Gamma_7$ and the second
two are for the $\Gamma_8$ conduction electrons.}
\protect\label{nca}
\end{figure}

\newpage
\begin{figure}
\vskip 1.0cm
\protect\caption[Fig.1.]
{{\bf Channel dependence of the Entropy.}
The NCA calculation of entropy clearly shows the right magnitude of
residual entropy depending on the relevant channel numbers. Solid
lines are Bethe-Ansatz curves. Note that the temperature scale is
linear for $M=1$. Referring to Table I: (a) $M=1$ case --
$\Box$ for model set 8; $\bigcirc$ for model set 7;
$\triangle$ for model set 6.
(b) $M=2$ case -- $\Box$ for model set 1; $\bigcirc$ for model set 2;
$\triangle$ for model set 3.
(c) $M=3$ case -- $\Box$ for model set 4; $\bigcirc$ for model set 5.}
\protect\label{entropy}
\end{figure}

\newpage
\begin{figure}
\vskip 1.0cm
\protect\caption[Fig.1.]
{{\bf Channel dependence of the Specific heat.} Solid lines are
Bethe-Ansatz results. The comparison to the Bethe Ansatz is
complicated by the background derived from the inter-configuration
peak. In the $M=1$ cases, the Kondo temperatures are adjusted
such that the NCA numerical results fall on the exact Bethe
Ansatz one. Thoughout all the figures presented in this paper
the estimated Kondo temperatures in the one-channel models
are used except for the
magnetic susceptibility curves (see Fig.~\ref{chi}).
Symbols have the same meaning as in Fig. \ref{entropy}.}
\protect\label{spheat}
\end{figure}

\newpage
\begin{figure}
\vskip 1.0cm
\protect\caption[Fig.1.]
{{\bf The scaling behavior of the static magnetic susceptibility.}
The static magnetic susceptibility obeys a scaling behavior for each
parameter regime leading to the $M=1,2,3$ fixed points.
The agreement with the Bethe ansatz results (solid lines) is good for
the overscreened cases ($M=2,3$).
$T_0=T_K/0.3$ for $M=2$, $T_K$ from Ref. \cite{sacramento}).
For convenience, we multiply $\chi(T)$ by 2.0 for $M=2$.
$M=1$ case -- $\Diamond$ for model set 8; $\star$ for model set 7;
$\ast$ for model set 6;
$M=2$ case -- $\Box$ for model set 1; $\bigcirc$ for model set 2;
$\triangle$ for model set 3;
$M=3$ case -- $+$ for model set 4; $\times$ for model set 5.
The ground state phase diagram for the model in
exchange coupling constant
parameter space is drawn in the inset, where
$g_i =  N(0)J_i$, $N(0)$ being the conduction band density of
states at the Fermi energy. The solid diagonal
line is for $M=3$.}
\protect\label{chi}
\end{figure}

\newpage
\begin{figure}
\vskip 1.0cm
\protect\caption[Fig.1.]
{{\bf Atomic spectral functions in the one-channel regime.}
$\rho_{01}$ is the interconfiguration spectral function which is
obtained from the convolution between $f^0$ and $f^1 \Gamma_7$ states.
$\rho_{12}$ is the interconfiguration spectral function which is
obtained from the convolution between $f^1 \Gamma_7$ and
$f^2 \Gamma_3$ states. One-channel Kondo effect leads to the
Kondo resonance development in $\rho_{01}$ just above the
Fermi level and the spectral depletion in $\rho_{12}$ right at
$\omega =0$. Spectral functions are displayed for model set 8.
The temperature variations are $T/D=3.678\times 10^{-2}$,
$1.077\times 10^{-2}$,
$3.155\times 10^{-3}$, $9.239\times 10^{-4}$,
$2.706\times 10^{-4}$, $7.924\times 10^{-5}$,
$2.321\times 10^{-5}$.}
\protect\label{rho1ch}
\end{figure}

\newpage
\begin{figure}
\vskip 1.0cm
\protect\caption[Fig.1.]
{{\bf Atomic spectral functions in the two-channel regime.}
Two-channel Kondo effect leads to the
Kondo resonance development in $\rho_{12}$ at the
Fermi level ($T=0$) and the spectral depletion in $\rho_{01}$ right at
$\omega =0$. Spectral functions are displayed for model set 1.
The temperature variations are the same as in Fig.~\ref{rho1ch}.}
\protect\label{rho2ch}
\end{figure}

\newpage
\begin{figure}
\vskip 1.0cm
\protect\caption[Fig.1.]
{{\bf Atomic spectral functions in the three-channel regime.}
For this parameter regime (model set 4),
two spectral functions are equivalent
in the asymptotic limit after a particle-hole transformation.
The temperature variations are the same as in Fig.~\ref{rho1ch}.}
\protect\label{rho3ch}
\end{figure}

\newpage
\begin{figure}
\vskip 1.0cm
\protect\caption[Fig.1.]
{{\bf Temperature dependence of the Kondo resonance peak.}
 The Kondo resonance peak position ($\omega_K$)
shows a different temperature dependence
on the relevant channel numbers.
(a) One-channel case ($\rho_{01}$): $\omega_K$ decreases
and saturates to a constant with decreasing temperature.
(b) Two-channel case ($\rho_{12}$) and (c) Three-channel case
($\rho_{12}$): $\omega_K$ decreases and tends to zero with
decreasing temperature. Symbols have the same meaning as
in Fig. \ref{entropy}.
Note the different temperature ranges between $M=1$ case
(top) and $M=2,3$ cases (bottom).}
\protect\label{kondopeak}
\end{figure}

\newpage
\begin{figure}
\vskip 1.0cm
\protect\caption[Fig.1.]
{{\bf Total atomic spectral function.}
The high energy structure essentially does not depend on either the temperature
or the relevant channel numbers. On the other hand, the relevant channel
number dependence shows up in the magnitude of the Kondo resonance structure.
(a) One-channel case (model set 8):
the Kondo resonance peaks are cut off to be compared with
the two- and three-channel cases.
(b) Two-channel case (model set 1). (c) Three-channel case (model set 4).
The temperature variations are the same as in Fig.~\ref{rho1ch}.}
\protect\label{rhof}
\end{figure}

\newpage
\begin{figure}
\vskip 1.0cm
\protect\caption[Fig.1.]
{{\bf Variation of $\chi^{''}(\omega)$ with temperature.}
The reduced dynamic magnetic susceptibility
$\tilde{\chi}^{''} (\omega,T) / \tilde{\chi}^{''} (\Gamma(T),T)$
is displayed and show a rough scaling behavior between two
exrema. One-channel case: model set 8; two-channel case: model set 1;
three-channel case: model set 4.
The temperature variations are the same as in Fig.~\ref{rho1ch}.}
\protect\label{dchi}
\end{figure}

\newpage
\begin{figure}
\vskip 1.0cm
\protect\caption[Fig.1.]
{{\bf Peak position ($\Gamma (T)$) of the dynamic magnetic susceptibility.}
The temperature variation of $\Gamma (T)$ depends on the relevant channel
numbers.
In the one-channel case (a), $\Gamma (T)$ approaches a constant value as
$T\to 0$. In the two- or three-channel cases (b) and (c),
$\Gamma (T\to 0) \to 0$
which is none other than Marginal Fermi liquid behavior.
Symbols have the same meaning as in Fig. \ref{entropy}.
Note the different temperature ranges between $M=1$ case
(top) and $M=2,3$ cases (bottom).}
\protect\label{gamma}
\end{figure}

\newpage
\begin{figure}
\vskip 1.0cm
\protect\caption[Fig.1.]
{{\bf Angle-averaged scattering rate.}
(a) One-channel case (model set 8): electron scattering is dominant.
The cusp feature for $M=1$ case right at the Fermi level
derives from the spectral depletion of $\rho_{12}$ and zeros
of the crystal harmonic $\Theta_{7}^{(5/2)} (\hat{k})$ which is
defined in the text.
(b) Two-channel case (model set 1): hole scattering is dominant.
(c) Three-channel case (model set 4): weak hole scattering
dominance over the
electron scattering comes from the degeneracy imbalance between
the $f^0$ singlet and $f^2 \Gamma_3$ doublet.
The temperature variations are the same as in Fig.~\ref{rho1ch}.}
\protect\label{tau}
\end{figure}

\newpage
\begin{figure}
\vskip 1.0cm
\protect\caption[Fig.1.]
{{\bf Low temperature dependence of resistivity.}
Correct power laws are found asymptotically
for the finite temperature NCA calculation in the two- and
three-channel cases. A power law of $T^{1/2} (T^{2/5})$ is
expected for $M=2(M=3)$.
Symbols have the same meaning as in Fig. \ref{entropy}.}
\protect\label{res23ch}
\end{figure}

\newpage
\begin{figure}
\vskip 1.0cm
\protect\caption[Fig.1.]
{{\bf Channel dependence of the thermopower.}
The thermopower is a measure of asymmetry in the density of
states and the scattering rate with respect to the Fermi level.
Large thermopower derives from the Kondo resonance scattering.
For $M=1$, the holes are main carriers leading to positive
thermopower. For $M=2,3$, electrons are dominant carriers
leading to a negative thermopower. The thermopower for
$M=2$ compares favorably with the experiments for the
stoichimetric system, CeCu$_2$Si$_2$. For more details,
see the text. }
\protect\label{tep}
\end{figure}

\newpage
\begin{table}
{Model parameters for the Ce impurity.
This set of model parameters covers the single, three, and
two channel Kondo regimes. $\Gamma=\Gamma_{01}=\Gamma_{12}$ is
the hybridization strength for both $f^0 - f^1$ and $f^1 - f^2$ mixing,
respectively. We scan channel number $M=1,2,3$ according to whether
$\epsilon_2>0,<0,=0$.
The Kondo scale $T_0$ is estimated as described in the text. }
\label{modelpar}
\begin{center}
\begin{tabular}{|c||c|c|c|c|c|} \hline
Set  & $M$ & $\Gamma/D$  &  $\epsilon_1/D$ &
 $\epsilon_2/D$ & $k_BT_0/D$ \\ \hline \hline
 1 & 2 &0.2  & $-0.4$ & $-0.1$ & $1.9081\times 10^{-3}$ \\
 2 & 2 & 0.2 & $-0.37$ & $-0.07$ & $1.5964\times 10^{-3}$ \\
 3 & 2 &0.2  &  $-0.35$ & $-0.05$ & $1.3492\times 10^{-3}$ \\
\hline
 4 & 3 & 0.2 & $-0.3$ & 0.0 & $1.5224\times 10^{-3}$ \\
 5 & 3 & 0.2 & $-0.4$ & 0.0 & $6.9413\times 10^{-3}$\\
\hline
 6 & 1  & 0.2 & $-0.3$ & 0.05 & $1.1957\times 10^{-3}$ \\
 7 & 1 & 0.2 & $-0.3$ & 0.07 & $1.6740\times 10^{-3}$ \\
 8 & 1  & 0.2 & $-0.3$ & 0.10 & $2.3914\times 10^{-3}$ \\
\hline
\end{tabular}
\end{center}
\end{table}

\end{document}